\documentclass[acmsmall,screen]{acmart}
\AtBeginDocument{%
  \providecommand\BibTeX{{%
    \normalfont B\kern-0.5em{\scshape i\kern-0.25em b}\kern-0.8em\TeX}}}

\newcommand{\Model}{\texttt{SeMPL}} % name of the model

\usepackage[ruled, linesnumbered, vlined, commentsnumbered]{algorithm2e}
\usepackage{amsmath,blkarray,booktabs}
\usepackage{xcolor,colortbl}
\usepackage{multirow}
\usepackage{subcaption} % for subfigure
\usepackage{tcolorbox}
\usepackage{csquotes}
\usepackage{tikz,pgfplots}
\usepackage{standalone}
\usepackage{pgfplots}
\usepackage{graphicx}
\usepackage{balance}
\usepackage{pifont}
\usepackage[english]{babel}
\definecolor{beaublue}{rgb}{0.74, 0.83, 0.9}
\newtcolorbox{quotebox}{colback=steel!10,boxrule=0.4pt,colframe=black,fonttitle=\bfseries,top=2pt,bottom=2pt}

\DeclareRobustCommand\dashed{\tikz[baseline=-0.6ex]\draw[thick,dashed] (0,0)--(0.54,0);}

\usepgfplotslibrary{fillbetween}

\definecolor{steel}{rgb}{0, 0.2, 0.9} 
\def\subfigsize{0.245}

\newcommand{\fixhd}[1]{%
  \smash[#1]{\vphantom{\Big|}}%
}

\newsavebox\mybox

\newtheorem{property}{Property}

\newcolumntype{P}[1]{>{\centering\arraybackslash}p{#1}}

\DeclareMathAlphabet\mathbfcal{OMS}{cmsy}{b}{n}
\SetKwInput{kwDeclare}{Declare}

\setcopyright{rightsretained}
\acmDOI{10.1145/3643743}
\acmYear{2024}
\copyrightyear{2024}
\acmSubmissionID{fse24main-p255-p}
\acmJournal{PACMSE}
\acmVolume{1}
\acmNumber{FSE}
\acmArticle{17}
\acmMonth{7}
\received{2023-09-28}
\received[accepted]{2024-01-23}
\begin{document}

%%
%% The "title" command has an optional parameter,
%% allowing the author to define a "short title" to be used in page headers.
%\title{Learning One at a Time: Predicting Configuration Performance in Multiple Environments with Sequential Meta-Learning}

\title[Predicting Configuration Performance with \Model]{Predicting Configuration Performance in Multiple Environments with Sequential Meta-Learning}

%\title{Learning One at a Time: Multi-Environment Prediction of Configuration Performance with Sequential Meta-Learning}

%%
%% The "author" command and its associated commands are used to define
%% the authors and their affiliations.
%% Of note is the shared affiliation of the first two authors, and the
%% "authornote" and "authornotemark" commands
%% used to denote shared contribution to the research.
\author{Jingzhi Gong}
% \authornote{This research was conducted when Jingzhi Going visited the University of Electronic Science and Technology of China.}
% %\authornote{Both authors contributed equally to this research.}
% %University of Electronic Science and Technology of China Chengdu, China
% \affiliation{%
%   \institution{University of Electronic Science and Technology of China}
%   % \city{Chengdu\\}
%   % \country{China}
%   \city{Chengdu}
%   \country{China}
%  }
 \affiliation{%
  \institution{Loughborough University}
  % \city{Chengdu\\}
  % \country{China}
  \city{Loughborough}
  \country{United Kingdom}
 }
% \email{jingzhigong0330@gmail.com}
\email{j.gong@lboro.ac.uk}

\author{Tao Chen}
\authornote{Tao Chen is the corresponding author (t.chen@bham.ac.uk).}
\affiliation{
  \institution{University of Birmingham}
  \city{Birmingham}
  \country{United Kingdom}
}
\email{t.chen@bham.ac.uk}

%%
%% The abstract is a short summary of the work to be presented in the
%% article.
\begin{abstract}

%such as performance testing, configuration tuning, and even runtime self-adaptation

%current work often builds performance models under a single environment, hence ignoring the rich knowledge from different settings and also restricting the accuracy for a new environment.

Learning and predicting the performance of given software configurations are of high importance to many software engineering activities. While configurable software systems will almost certainly face diverse running environments (e.g., version, hardware, and workload), current work often either builds performance models under a single environment or fails to properly handle data from diverse settings, hence restricting their accuracy for new environments. In this paper, we target configuration performance learning under multiple environments. We do so by designing \Model---a meta-learning framework that learns the common understanding from configurations measured in distinct (meta) environments and generalizes them to the unforeseen, target environment. What makes it unique is that unlike common meta-learning frameworks (e.g., \texttt{MAML} and \texttt{MetaSGD}) that train the meta environments in parallel, we train them sequentially, one at a time. The order of training naturally allows discriminating the contributions among meta environments in the meta-model built, which fits better with the characteristic of configuration data that is known to dramatically differ between different environments. Through comparing with 15 state-of-the-art models under nine systems, our extensive experimental results demonstrate that \Model~performs considerably better on $89\%$ of the systems with up to $99\%$ accuracy improvement, while being data-efficient, leading to a maximum of $3.86\times$ speedup. All code and data can be found at our repository: \texttt{\textcolor{blue}{\url{https://github.com/ideas-labo/SeMPL}}}.
\end{abstract}



%%
%% Keywords. The author(s) should pick words that accurately describe
%% the work being presented. Separate the keywords with commas.

%%
%% This command processes the author and affiliation and title
%% information and builds the first part of the formatted document.
\maketitle

\section{Introduction}
\label{sec:introduction}

Most software systems can be flexibly configured to meet the needs of a certain scenario. This is achieved by jointly adjusting various configuration options, which can determine, e.g., the size of the threads pool; the capacity of the cache; or the underlying algorithm to use~\cite{DBLP:journals/tse/SayaghKAP20,10.1145/3514233,DBLP:journals/tosem/ChenL23a}. As such, the configuration will inevitably lead to considerable impacts on the performance, e.g., runtime and throughput. Even carefully managed configuration could still be error-prone and result in devastating consequences---a previous study finds that 59\% of the configuration-related issues have caused severe performance concerns in modern configurable software systems~\cite{DBLP:conf/esem/HanY16}.

The key to configuration management is how to accurately infer the performance of the given configurations via a performance model, serving as the foundation for, e.g., performance debugging~\cite{DBLP:conf/eurosys/IqbalKJRJ22}, configuration tuning~\cite{DBLP:journals/tse/Nair0MSA20}, and even self-adaptation~\cite{DBLP:conf/wcre/Chen22,DBLP:journals/tsc/ChenB17,DBLP:journals/tosem/ChenLBY18}. This, however, can neither be realistically tackled by analytical models (due to diverse types of configuration options~\cite{DBLP:conf/sigsoft/0001L21}) nor profiling (due to expensive measurement~\cite{DBLP:journals/tse/Nair0MSA20}). Through learning on the available data, current work has successfully leveraged machine learning to build various performance models~\cite{DBLP:conf/msr/GongChen22}. Yet, those approaches mostly focus on configuration performance learning under one environment~\cite{DBLP:conf/sigsoft/GongChen2023,DBLP:conf/icse/SiegmundKKABRS12,DBLP:journals/ese/GuoYSASVCWY18,DBLP:conf/icse/HaZ19,DBLP:conf/splc/ValovGC15,DBLP:conf/oopsla/QueirozBC16,DBLP:conf/icse/0003XC021,DBLP:journals/sqj/SiegmundRKKAS12}, e.g., a pre-defined workload, a fixed hardware, and a specific version.

%At ICSE'23, there is a paper~\cite{muhlbaueranalyzing} that investigates the impacts of workloads on the software configuration performance, e.g., runtime, and how these might change the design of a machine learning-based performance model. The authors concluded that:

Working on a single environment is an over-optimistic assumption as it is not uncommon to see configurable software systems run under diverse conditions. For example, a database system may experience both read-heavy and write-heavy workload~\cite{DBLP:conf/kbse/JamshidiSVKPA17}. Similarly, the hardware between the testing and production infrastructure might be drastically different~\cite{DBLP:journals/toit/LeitnerC16,DBLP:journals/corr/BrunnertHWDHHHJ15}, especially during the modern DevOps era. The ignorance of multiple environments would inevitably harm the effectiveness of a performance model.~\citet{DBLP:conf/kbse/JamshidiSVKPA17} reveal that the accuracy of a single environment model can be severely degraded when used in a different environment. Furthermore, due to the expensive measurement of configuration, e.g., it can take hours or days to measure only a few configurations~\cite{DBLP:conf/wosp/ValovPGFC17,DBLP:conf/sigsoft/0001L21}, building a new model for every distinct environment is unrealistic. Recently, at ICSE'23,~\citet{muhlbaueranalyzing} have demonstrated that predicting under multiple environments can pose significant threats to the robustness and generalizability of performance models learned using a single environment. Through a large-scale study, they concluded that:

\begin{quote}
\textit{``Performance models based on a single workload are useless, unless the configuration options’ sensitivity to workloads is accounted for.''}
\end{quote}

As a result, the failure to take multiple environments into account when learning the performance model for configurable software not only degrades accuracy but also incurs extra overhead of model re-building, as the valuable data samples measured under different environments are wasted~\cite{DBLP:journals/corr/abs-1911-01817, DBLP:conf/cloudcom/IorioHTA19, DBLP:journals/corr/abs-1803-03900}. This leads to a previously unaddressed problem: \textit{How to effectively leverage configurations measured in different environments for modeling configuration performance?} Yet, learning a performance model under multiple environments for configurable software is challenging due to the large variations between data measured in distinct environments. For example, several studies have revealed that varying environments can cause substantial changes in the performance distributions with non-monotonic correlations, including workloads~\cite{muhlbaueranalyzing,DBLP:conf/wcre/Chen22,DBLP:conf/wosp/PereiraA0J20}, versions~\cite{DBLP:journals/tse/MartinAPLJK22} and hardware~\cite{DBLP:conf/kbse/JamshidiSVKPA17}.

%Yet in another example, Chen~\cite{DBLP:conf/wcre/Chen22} discovered that the workloads significantly alter the multi-modality of configuration landscapes. 

%  what is the finding of that work? large performance vairation

Despite being uncommon,~\citet{muhlbaueranalyzing} summarized two major categories from existing work on how multiple environments have been handled, each with their own limitations:

\begin{enumerate}
  
    \item \textbf{Environment as additional features~\cite{DBLP:journals/tse/ChenB17,DBLP:conf/kbse/KocMWFP21,DBLP:conf/europar/LengauerABGHKRTGKKRS14,DBLP:conf/icse/Chen19b}:} Here, the specific properties of an environment, e.g., size and job counts, are considered as model features alongside the configuration options in performance model learning. However, in this category, not only the additional measurements can be costly, but the extra dimension(s) also make the true concept more difficult to learn and generalize.

%~\cite{DBLP:conf/icse/ShariflooMQBP16, DBLP:conf/icse/JamshidiVKSK17, DBLP:conf/kbse/JamshidiSVKPA17, DBLP:journals/sosym/KolesnikovSKGA19,DBLP:journals/corr/abs-1911-01817,DBLP:journals/tse/MartinAPLJK22,DBLP:conf/wosp/ValovPGFC17}    

    \item \textbf{Transfer learning~\cite{ DBLP:conf/icse/JamshidiVKSK17,DBLP:journals/corr/abs-1911-01817,DBLP:journals/tse/MartinAPLJK22,DBLP:conf/wosp/ValovPGFC17}:} Given an existing performance model trained in the source environment(s), its differences to the new environment can be learned via transfer learning models. Yet, the key shortcoming thereof is that the loss function in transfer learning is tailored to a specific target environment, hence lacking generalizability to arbitrary environments.
    
    %it requires knowing a good amount of samples from the target environment to be a major part of the training.

\end{enumerate}

In this paper, we propose the \textit{third category} to address the aforementioned gap and challenges using the concept of meta-learning---a form of machine learning that is capable of ``learning to learn'' by learning data across multiple environments (or tasks\footnote{In this paper, we use task and environment interchangeably.}), and generalize the learning to an unforeseen one. From this, we present \underline{\textbf{Se}}quential \underline{\textbf{M}}eta \underline{\textbf{P}}erformance \underline{\textbf{L}}earning (\Model), a framework that produces a general meta-model learned from data of all existing meta environments during pre-training. When needed, such a meta-model can then be quickly adapted for accurately predicting performance under a target environment via fine-tuning. What makes \Model~unique is that, unlike popular general-purpose meta-learning frameworks such as \texttt{MAML}~\cite{DBLP:conf/icml/FinnAL17} and \texttt{MetaSGD}~\cite{DBLP:journals/corr/LiZCL17} which learn the data of meta environments in parallel, \Model~follows sequential meta-learning that learns the datasets of meta environments one at a time in a specific order, aiming to discriminate their contributions in the meta-model built. This is a tailored design to deal with the potentially large variations between measurements in different environments from the configuration data. 

Specifically, our contributions are:

\begin{itemize}
    \item We justify, both analytically and empirically, why the concept of sequential meta-learning in \Model~is more suitable for learning multi-environment configuration data than general frameworks such as \texttt{MAML} and \texttt{MetaSGD}, according to three unique properties in \Model:

    \begin{enumerate}
        \item the sequence matters;
        \item train later contributes more;
        \item and using more meta environments are beneficial.
    \end{enumerate}

    \item Drawing on those properties provided by \Model, we design a novel method that selects the optimal sequence of learning the data of meta environments, which fits the characteristics of the configuration data under multiple environments.

    \item Based on nine systems with 3-10 meta environments and five training data sizes each, we experimentally compare \Model~against 15 state-of-the-art models for single or multiple environments, taken from the software, system, and machine learning communities.

    %\item Through 9 systems of diverse domains, 3-10 different target tasks, and 5 distinct training data sizes each, we experimentally compare \Model~with 4 single-environmental models; 5 models that consider meta-tasks in random orders, and 4 state-of-the-art transfer learning and meta-learning models, including those from both performance modeling and the general machine learning community.
\end{itemize}

The results are encouraging: \Model~significantly outperforms existing models in $89\%$ of the systems with up to $99\%$ accuracy improvement; it is also data-efficient with at most $3.86\times$ speedup.

%the single environment models; dramatically improves the accuracy of models that handle multiple environments; can be greatly benefited from the sequence selection; can be improved quickly as long as some data is available for the pre-training.

The paper is organized as follows: Section~\ref{sec:background} overviews the preliminaries and related work. Section~\ref{sec:premises} elaborates the theory behind this work and Section~\ref{sec:framework} illustrates the algorithm and details of the \Model~framework. Section~\ref{sec:setup} presents the experimental setup followed by the experiment results in Section~\ref{sec:evaluation}. Discussion and conclusion are presented in Section~\ref{sec:discussion} and~\ref{sec:conclusion}, respectively.

\section{Preliminaries and Related Work}
\label{sec:background}

%In this section, we discuss the necessary background and the related work for this research.

\subsection{Single Environment Configuration Performance Learning}
\label{sec:bg-single}
Learning performance for configurable software systems is commonly formulated as single-task learning that learns and generalizes data under a single environment. Formally, we aim to build a regression model $f$ that predicts the performance $p$ of an unforeseen configuration $\mathbf{\overline{x}'}$:
\begin{equation}
\begin{aligned}
 &\text{Train: } \mathbfcal{E}_{target} \Longrightarrow f\\
 &\text{Predict: } f(\mathbf{\overline{x}'})=p_{target} \text{ | } \mathbf{\overline{x}'}\in \mathbfcal{E}_{target}
\end{aligned}
    \label{eq:prob}
\end{equation}
whereby $\mathbfcal{E}_{target}$ denotes the training samples of configuration-performance pairs ($\{\mathcal{C} \rightarrow \mathcal{P}\}$) measured under the target environment, such that $\mathbf{\overline{x}} \in \mathbfcal{E}_{target}=\{\mathcal{C} \rightarrow \mathcal{P}\}$. $\mathbf{\overline{x}}$ is a measured configuration and $\mathbf{\overline{x}}=(x_{1},x_{2},\cdots,x_{n})$, where there are $n$ configuration options and each option $x_{i}$ can be either binary or categorical/numerical. $f$ should be trained in such a way that its prediction ($p_{target}$) on $\mathbf{\overline{x}'}$ is as close to the actual performance as possible. Many models for learning configuration performance in a single environment have been proposed~\cite{DBLP:conf/sigsoft/GongChen2023,DBLP:conf/icse/SiegmundKKABRS12,DBLP:journals/ese/GuoYSASVCWY18,DBLP:conf/icse/HaZ19,DBLP:conf/splc/ValovGC15,DBLP:conf/oopsla/QueirozBC16,DBLP:conf/icse/0003XC021,DBLP:journals/sqj/SiegmundRKKAS12}. Here, we briefly explain some state-of-the-art models, which are also compared in Section~\ref{sec:evaluation}:

%The corresponding performance is denoted as $\mathcal{P}$.

%including linear models[], tree-liked models[], neural networks[], and ensemble models[]. 

\begin{itemize}
    \item \textbf{\texttt{DeepPerf}}~\cite{DBLP:conf/icse/HaZ19}: a Deep Neural Network (DNN) model with $L_1$ regularization and fast hyperparameter tuning for sparse performance learning. 
    %It has been compared as a SOTA performance model in most of the recent studies since 2019~\citep{DBLP:conf/esem/ShuS0X20}.
    \item \textbf{\texttt{DECART}}~\cite{DBLP:journals/ese/GuoYSASVCWY18}: an improved regression tree model~\cite{breiman2017classification} with a specialized sampling mechanism.
    \item \textbf{\texttt{Random Forest} (\texttt{RF})}~\cite{DBLP:conf/splc/ValovGC15,DBLP:conf/oopsla/QueirozBC16,DBLP:conf/icse/0003XC021}: an ensemble of trees that tackle the feature sparsity issue.
    \item \textbf{\texttt{SPLConqueror}}~\cite{DBLP:journals/sqj/SiegmundRKKAS12}: a baseline model that relies on linear regression.
    \item \textbf{\texttt{XGBoost}}~\cite{DBLP:conf/kdd/ChenG16}: a gradient boosting algorithm that leverages the combination of multiple trees to create a robust ensemble.
\end{itemize}

%Yet, when the environment under which the software system operates changes, a model will often need to be retained [].

\subsection{Configuration Performance Learning with Multiple Environment Inputs}
\label{sec:bg-env}

%When running under diverse environments (e.g., versions, hardware, and workloads), [] have shown that configurable software systems can exhibit rather distinct characteristics in the configuration data. [] also demonstrate that varying workloads can create a large variation in performance even given the same configuration. Therefore, in such a case, models learned in a single environment will inevitably be less effective without special treatment. 

%In the presence of diverse environments (e.g., versions, hardware, and workloads), models learned in a single environment will inevitably be less effective without special treatment due to the dramatically changing characteristics in the configuration data~\cite{DBLP:conf/kbse/JamshidiSVKPA17}. 

To handle multiple environments, a natural way is to combine the environmental features with configuration options in the model. In this way, Equation (1) remains unchanged but the configuration-performance pairs from all available environments are merged and the environment features serve as additional inputs, denoted as $\mathcal{E}$, i.e., $\mathbfcal{E}_{target}=\{\mathcal{C} \times \mathcal{E} \rightarrow \mathcal{P}\}$. For example,~\citet{DBLP:conf/icse/Chen19b} has followed this strategy to examine various types of underlying models. The environmental features therein are the workload and request rate, \textit{etc}. More domain-specific environmental features also exist, such as those for program verification~\cite{DBLP:conf/kbse/KocMWFP21}, high-performance computing~\cite{DBLP:conf/europar/LengauerABGHKRTGKKRS14}, and cloud computing~\cite{DBLP:journals/tse/ChenB17}.

However, a major limitation thereof is that the combined input space of options and environmental features makes the training even more difficult, as not all options are environment-sensitive~\cite{LESOIL2023111671}.

%and it also exacerbates the issue of large variation between data in different environments, 

%\subsection{Learning Performance in Multiple Environments}
%\subsection{Transfer and Multi-Task Learning for Performance Models}
\subsection{Joint Learning for Configuration Performance}
\label{sec:bg-transfer}

\subsubsection{Transfer Configuration Performance Learning}

Building performance models under multiple environments can be also handled by transfer learning, in which Equation (1) is changed to:
\begin{equation}
\begin{aligned}
    &\text{Train: } \mathbfcal{E}_{1}\cup\mathbfcal{E}_{2}...\cup\mathbfcal{E}_{m}\cup\mathbfcal{E}_{target} \Longrightarrow f\\
    &\text{Predict: } f(\mathbf{\overline{x}'})=p_{target} \text{ | } \mathbf{\overline{x}'}\in \mathbfcal{E}_{target}
\end{aligned}
    \label{eq:prob1}
\end{equation}
Here, by treating the learning of data under different environments as independent tasks, both the data from source environment(s) ($\mathbfcal{E}_{1},...,\mathbfcal{E}_{m}$) and target environment ($\mathbfcal{E}_{target}$) are jointly learned by a base learner with information sharing on, e.g., model parameters/data samples~\cite{ DBLP:conf/icse/JamshidiVKSK17,DBLP:journals/corr/abs-1911-01817}, features~\cite{DBLP:journals/tse/MartinAPLJK22}, or prediction outputs~\cite{DBLP:conf/wosp/ValovPGFC17}. Among others, \textbf{\texttt{BEETLE}}~\citep{DBLP:journals/corr/abs-1911-01817} is a transfer learning model for configuration performance learning. The key idea is to evaluate all the source-target pairs over the candidate environments and identify the best source as the ``bellwether'' to use. The data of the source and target environment are jointly learned by a regression tree model. \textbf{\texttt{tEAMS}}~\citep{DBLP:journals/tse/MartinAPLJK22} is a recent model caters for the evolution of configurable software. With feature alignment, the model trained on the data of an older version is transferred to learn the data under the new version.

Yet, the key limitation of transfer learning is that, regardless of how many environments serve as the source data, the loss function therein needs to be specific to the target environment, making it difficult to generalize to the other environments~\cite{torrey2010transfer}.

%it often only learns data from one source environment ($\mathbfcal{E}_{source}$)~\cite{torrey2010transfer}, hence wasting the knowledge that could have been extracted from the readily available samples under other environments

%\footnote{Indeed, multiple candidate sources may be considered in transfer learning but often only the best one is selected for learning, such as in \texttt{BEETLE}~\citep{DBLP:journals/corr/abs-1911-01817}.}. 

\subsubsection{Multi-Task Configuration Performance Learning}

%Specifically, with multi-task learning, the formalization becomes:

To cope with the above limitation, a relevant paradigm is multi-task learning, which can also learn data from multiple environments as the sources ($\mathbfcal{E}_{1},...,\mathbfcal{E}_{m}$)~\cite{DBLP:conf/iclr/YangH17,DBLP:conf/eurosys/AlabedY21,DBLP:conf/socpros/MadhumathiS18} and has the ability to generalize the predictions for all environments learned simultaneously. Here, Equation (2) will be changed to:
\begin{equation}
\begin{aligned}
    &\text{Train: } \mathbfcal{E}_{1}\cup\mathbfcal{E}_{2}...\cup\mathbfcal{E}_{m}\cup\mathbfcal{E}_{target} \Longrightarrow f\\
    &\text{Predict: } f(\mathbf{\overline{x}'})=\{p_1,p_2,...,p_m,p_{target}\} \text{ | } \mathbf{\overline{x}'}\in \mathbfcal{E}_{target}
\end{aligned}
\label{eq:prob2}
\end{equation}
whereby $p_m$ is the predicted performance of $\mathbf{\overline{x}'}$ under the $m$th environment. There is no clear distinction between source and target environments, as the key in multi-task learning is to share information on data from whatever available environments. For example,~\citet{DBLP:conf/socpros/MadhumathiS18} propose \textbf{\texttt{Multi-Output Random Forest} (\texttt{MORF})}, a multi-task learning version of Random Forest where there is one dedicated output for each environment of performance prediction. 

%The trained model can be directly used when any of the environments involved need prediction.

%Since multi-task learning models are rarely used for configuration performance learning, in this work, we examine the following representative, which has been used to model the performance of configurable networked systems:

However, the training requires foreseeing the target environment and suffers ``negative transfer'', as task gradients
may interfere, overcomplicating the loss function landscape~\cite{DBLP:conf/icml/StandleyZCGMS20}.

%a good amount of labeled data (measured configurations) thereof [], as it often needs balanced data samples from all tasks to jointly learn the relationships between them.

\subsection{Meta-Learning for Configuration Performance}
\label{sec:bg-meta}

Unlike transfer and multi-task learning, meta-learning seeks to learn a meta-model based on some readily available data from known environments (a.k.a meta environments) without foreseeing the target environment~\cite{DBLP:journals/air/VilaltaD02}. Formally, in meta-learning, the process becomes:
\begin{equation}
\begin{aligned}
    &\text{Train: } \mathbfcal{E}_{1}\cup\mathbfcal{E}_{2}...\cup\mathbfcal{E}_{m} \Longrightarrow f' \text{; } \mathbfcal{E}_{target}+f' \Longrightarrow f\\
   &\text{Predict: } f(\mathbf{\overline{x}'})=p_{target} \text{ | } \mathbf{\overline{x}'}\in \mathbfcal{E}_{target}
\end{aligned}
\label{eq:prob3}
\end{equation}
The key idea is that by learning the data of meta environments using a base learner at the meta-level, the meta-model $f'$ will obtain the ability of ``learning to learn'', enabling easier fine-tuning on the new target environment ($\mathbfcal{E}_{target}$) when it becomes available. As such, meta-learning achieves: (1) fast adaptation; and (2) a better generalized and more accurate model~\cite{DBLP:conf/icse/JamshidiVKSK17}.

% \begin{enumerate}
% \item being agnostic to the base learner;
% \item fast adaptation to the target-task;
% \item better generalization and hence a more accurate model.
% \end{enumerate}

%\subsubsection{Parallel Learning}

Although rarely used for configuration performance learning, there exist different models for meta-learning in the machine learning community~\cite{DBLP:conf/icml/FinnAL17,DBLP:journals/corr/LiZCL17,DBLP:conf/icml/YaoWHL19,DBLP:conf/nips/ChuaLL21}, in which the meta tasks are learned simultaneously and the outcomes are combined. The most noticeable ones are \texttt{MAML}~\cite{DBLP:conf/icml/FinnAL17} and its variant \texttt{MetaSGD}~\cite{DBLP:journals/corr/LiZCL17}: \textbf{\texttt{MAML}}~\citep{DBLP:conf/icml/FinnAL17} is a state-of-the-art meta-learning framework that has been widely applied in different domains~\citep{DBLP:conf/aistats/ChenC22}, including software engineering~\citep{DBLP:conf/icse/ChaiZSG22}. \texttt{MAML}, if used for configuration performance learning, learns a set of good model parameter values on data of the meta environments in parallel, building a meta-model ($f'$), which can then be adapted to build a fine-tuned model $f$ for the target environment. In contrast, \textbf{\texttt{MetaSGD}}~\citep{DBLP:journals/corr/LiZCL17} extends the \texttt{MAML} by additionally adapting the learning rate along the meta-training, expediting the learning over \texttt{MAML}.

% \begin{itemize}
%     \item \textbf{\texttt{MAML}}~\citep{DBLP:conf/icml/FinnAL17}: a state-of-the-art meta-learning framework that has been widely applied in different domains, including software engineering~\citep{DBLP:conf/aistats/ChenC22}. \texttt{MAML} learn a set of good model parameter values via training on the meta-tasks in parallel, building a meta-model ($f'$), which can then be adapted to fit with a target-task to build a fine-tuned model $f$. 
%     %but still effective meta learning model which trains meta tasks separately on the initial model and uses gradients of the resulting meta models to learn sensitive initial parameters for fast fine-tuning. MAML has been the state-of-the-art meta learning model in a wide range of fields since 2017
%     \item \textbf{\texttt{MetaSGD}}~\citep{DBLP:journals/corr/LiZCL17}: a model extended based on \texttt{MAML} by additionally adapting the learning rate along the pre-training process. In this way, it has been shown to achieve faster learning than \texttt{MAML}
%     %similar to MAML, seeks to use the meta tasks to learn an optimal initialization and a best learning rate, such that the trained meta model can fit the fine-tuning task faster and better. By updating the learning rate,~\citet{DBLP:journals/corr/LiZCL17} suggests the model has a higher capacity than MAML.
% \end{itemize}

Nevertheless, in Section~\ref{sec:premises}, we will explain why general models like \texttt{MAML} are relatively ill-suited to configuration performance learning and elaborate on the theory behind the proposed \Model~framework.

%%%%%%%%%%%%%%%

% \subsubsection{Sequential Learning}
% %For example, Zhang \textit{et al.}~\cite{DBLP:conf/sigir/ZhangFW00LZ20} retrain a neural network according to the sequence of arrived tasks in recommendation systems.
% Albeit that learning meta tasks simultaneously is predominated for meta-learning, there indeed exists work that takes the sequence of training meta tasks into account~\cite{DBLP:conf/sigir/ZhangFW00LZ20}. For example, Finn \textit{et al.}~\cite{pmlr-v97-finn19a} refine \texttt{MAML} to handle the meta tasks that occur in causally related rounds with an additional local update for each task.

% However, the notion of sequence in the above work is derived from strong causal relationships between the meta-tasks, e.g., the time-series occurrence of events. This, however, is often not the case for configuration performance learning. Yet, with the proposed \Model, we show that a certain sequence of learning the data of distinct meta environments can be particularly beneficial.

%For configuration performance learning, there may be no such casual correlations, as measurements under different environments can be conducted in any order. Y
%dramatically improve the effectiveness for predicting configuration performance.

\section{The Theory behind \Model~for Configuration Performance Learning}
\label{sec:premises}

\begin{figure}[!t]
\centering
    \begin{subfigure}[t]{0.43\columnwidth}
    \includegraphics[width=\columnwidth]{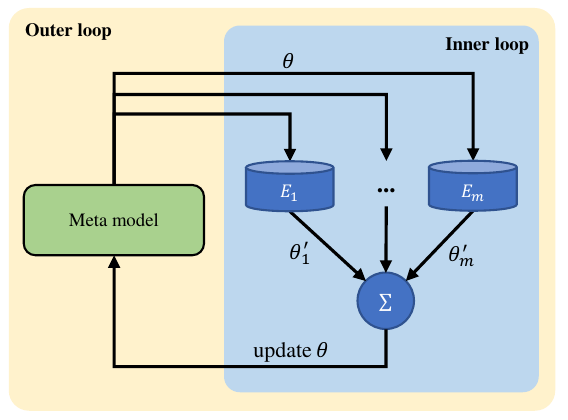}
    \subcaption{\texttt{MAML}}
    \label{subfig:maml model}
    \end{subfigure}
~\hspace{0.4cm}
%\hfill
    \begin{subfigure}[t]{0.43\columnwidth}
    \includegraphics[width=\columnwidth]{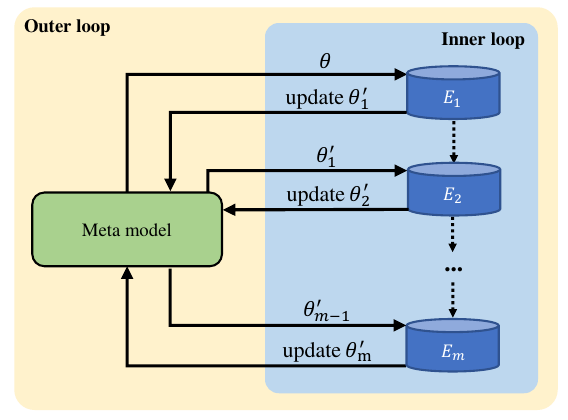}
    \subcaption{\Model}
    \label{subfig:sequential model}
    \end{subfigure}
    \vspace{-0.3cm}
\caption{Workflow of \texttt{MAML} and the proposed \Model. The meta-model can be produced by any base learner.}
 %\vspace{-0.4cm}
\label{fig:models}
\end{figure}

In essence, the key to the success of \texttt{MAML} (and its variants) is that it can produce a set of good model parameter values for the target environment to start training, paired with any gradient descent-based learner, which enables fast adaptation of the created meta-model~\cite{DBLP:conf/iclr/RaghuRBV20}. As in Figure~\ref{fig:models}a, this is achieved via iteratively updating the parameters of the meta-model ($\theta$ for the outer loop\footnote{The default iteration limit of the outer loop in \texttt{MAML} is 1.}) using those trained on the data of each meta environment individually (i.e., $\theta'_1,...,\theta'_m$ from the inner loop). In such a process, data of all meta environments are learned simultaneously followed by a linear aggregation of their model parameter values, hence they provide equal contributions to the meta-model. This, however, is ill-suited for the performance learning of configurable software systems. Because, unlike some other domains, prior studies have revealed that the data from distinct environments exhibit drastically different correlations between performance and configuration:

\begin{itemize}

\item A study on \textsc{x264}~\cite{DBLP:conf/wosp/PereiraA0J20} showed that different video inputs can lead to varying and non-monotonic performance results.

\item \citet{DBLP:conf/wcre/Chen22} discovered dramatic configuration landscape shifts due to workload changes.

\item \citet{muhlbaueranalyzing} reported that varying workloads can change how configuration options affect performance, causing drastic variations among performance distributions.

\item \citet{DBLP:conf/kbse/JamshidiSVKPA17} revealed the non-linear correlation on data from different workloads, software versions and hardware.
\end{itemize}

As a result, these prior observations derive the following insight: 

%for configuration performance learning with multiple environments:

\begin{quotebox}
   \noindent
   \textit{\textbf{Insight of Learning Configuration Data:}} The ideal parameter values and distributions for models learned under different environments can be, unfortunately, rather different. 
\end{quotebox}

This means that treating all meta environments equally in \texttt{MAML} can force some less useful (or even misleading) meta environments to contribute, causing some model parameter values to largely deviate from the optima required for the target environment. Therefore, what we need is a tailored model for multi-environment configuration performance learning with the following requirements:

\begin{itemize}
    \item \textbf{Requirement 1:} Configuration data of distinct meta environments should be discriminated.
    \item \textbf{Requirement 2:} Configuration data of distinct meta environments should have alterable contributions to the learning.
    \item \textbf{Requirement 3:} The valuable information of data from different meta environments should be exploited fully since measuring configurations is highly expensive~\cite{DBLP:conf/sigsoft/0001L21,DBLP:journals/tse/Nair0MSA20}.
\end{itemize}

%This is also one of the reasons that prevent the success of models based on additional environment inputs and multi-task learning.

A naive (perhaps natural) solution to the above is to consider weights between meta environments when updating the $\theta$ in \texttt{MAML}. However, this has the following shortcomings:

\begin{itemize}
\item It is difficult to precisely quantify the relative contributions between meta environments, as the model parameter values of neurons are naturally multi-dimensional. 

%For example, setting a meta-task with 2 and another with 1 does not necessarily ensure that the former is twice as important as the latter.

\item Setting/updating the weights is challenging, especially with more meta environments. 

\item It has been shown that the weights can obscure the optimization during the training~\cite{10.1145/3514233}.
\end{itemize}

Instead of weights, we propose sequential meta-learning in \Model: as in Figure~\ref{fig:models}b, the datasets in distinct meta environments are learned one by one in a certain order, and so does the update of meta-model's parameters $\theta$. The sequence of meta environments prioritizes their contributions to learning, thereby resolving the limitation of \texttt{MAML} for configuration performance learning without extra parameters. In what follows, we elaborate on the theory behind \Model~and its properties.
%why not weight
%breif introduction of the approach

\subsection{The Sequence Matters}

The very first property that we intentionally design for \Model, which meets \textbf{Requirement 1}, is:

\begin{property}
\label{pro:1}
The sequence of learning the data in distinct meta environments significantly influences the meta-model built, hence leading to different adaptation effects for the target environment.
\end{property}

%The heuristic of gradient decent in the training process would always tend to fit more for the meta-task trained later, hence the parameter values of the meta-model will have a large proportion of distribution close to that of single task training under $\mathbfcal{E}_{i+1}$ than that when training on $\mathbfcal{E}_i$ alone.

Given that the datasets in meta environments are learned in sequence with only one being considered each time, the model parameter values for a meta environment trained earlier would serve as the initial values of the ones that will be learned later. As such, different sequences would cause distinct orders of intermediate initialization on the model parameter values for learning some meta environments. Since the heuristic of gradient descent in training is often stochastic and the training sample size can be limited for configurable software systems, working on different sets of starting points would most likely have different results, leading to diverse meta-models in the end.

\begin{figure}[!t]
  \centering
   \begin{subfigure}[t]{0.75\columnwidth}
        \centering
\includegraphics[width=\textwidth]{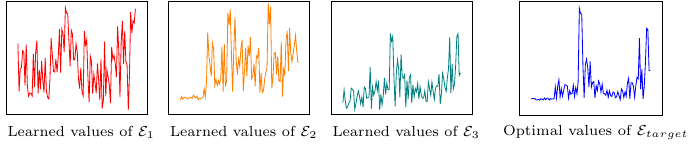}
        \subcaption{Configuration performance learning under independent single environment}
   \end{subfigure}

      \begin{subfigure}[t]{0.75\columnwidth}
        \centering
\includegraphics[width=\textwidth]{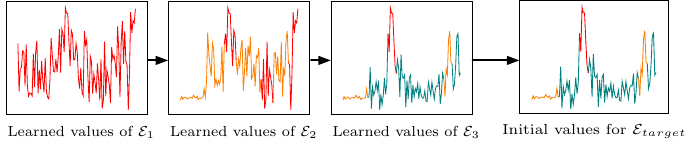}
      \subcaption{Multi-environment configuration performance learning with \Model}
   \end{subfigure} 
 
    \begin{subfigure}[t]{0.75\columnwidth}
        \centering
\includegraphics[width=\textwidth]{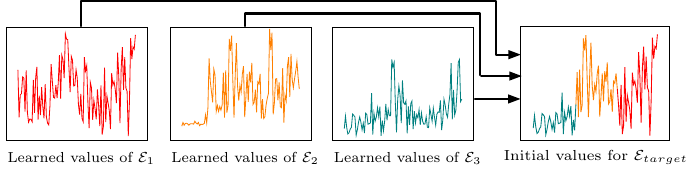}
      \subcaption{Multi-environment configuration performance learning with \texttt{MAML}}
   \end{subfigure}  
    
   %\vspace{-0.3cm}
 \caption{Illustrating the distributions of the model parameter values in different situations under a real-world software system; the base learner is a regularized Deep Neural Network (it is best viewed in color). The x- and y-axis are model parameters and their corresponding performance values, respectively.}
      \label{fig:theory}
     \vspace{-0.4cm}
  \end{figure}

For example, Figure~\ref{fig:theory}a illustrates the distributions of model parameter values when trained individually on data of three single environments, namely $\mathbfcal{E}_1$, $\mathbfcal{E}_2$, and $\mathbfcal{E}_3$. With \Model~in this case (Figure~\ref{fig:theory}b), the model parameter values trained from $\mathbfcal{E}_1$ becomes the initial values for learning the data of $\mathbfcal{E}_2$, after which the values become the initials for learning the $\mathbfcal{E}_3$ data. Here, the resulted model parameter values when learning the data of $\mathbfcal{E}_2$, whose initialization is set by learning the data of $\mathbfcal{E}_1$, would lead to a distribution mixed from those of learning the data of each meta environment individually (Figure~\ref{fig:theory}a). As such, if we swap $\mathbfcal{E}_1$ with $\mathbfcal{E}_3$, the resulting distribution of model parameter values would be different. \texttt{MAML}, in contrast, is insensitive to the sequence of meta environments as all of them are learned simultaneously and the learned model parameter values are aggregated to update the meta-model, as in Figure~\ref{fig:theory}c. Figure~\ref{fig:ranking_MRE}a shows the empirical results for a randomly selected system and we observed similar outcomes for the others. Clearly, the different sequences in the meta-learning of \Model~can lead to distinct accuracy for the target environment.

\subsection{Train Later Contributes More}

To satisfy \textbf{Requirement 2}, a related property that can be derived from \textbf{Property 1} in \Model~is:

\begin{property}
\label{pro:2}
If the data of meta environment $\mathbfcal{E}_{i+1}$ is learned later than that of $\mathbfcal{E}_i$, then the model parameter values of meta-model are more similar to those of learning $\mathbfcal{E}_{i+1}$ data alone than those trained under $\mathbfcal{E}_i$. Therefore, data of the more useful meta environment should be learned later.
\end{property}

With the sequential training of meta environments data in \Model, it is easy to see that, if there are $n$ meta environments ($n>0$), the model parameter values learned on $\mathbfcal{E}_i$ will experience $n-i$ ($i \in [1,n]$) updates later on. Since in each update, the training tends to tune the model parameter to fit the data in the corresponding meta environment, which can be rather different to $\mathbfcal{E}_i$ due to the characteristics of configuration landscape, the distribution of model parameter values learned from $\mathbfcal{E}_i$ will be gradually overridden with more subsequent updates. As a result, the bigger the $i$, i.e., a meta environment sits at a later position of the sequence, the fewer updates to the parameters trained under $\mathbfcal{E}_i$, allowing it to preserve more model parameter values (and distribution) in the meta-model and contribute more therein than the meta environments learned earlier. Importantly, this leads to an obvious rule: the more useful meta environments---those that can provide better initial model parameter values for the target environment---should be left to later positions (we will explain how we measure the usefulness in Section~\ref{sec:framework}).

\begin{figure}[!t]
\centering
\footnotesize
 \begin{subfigure}{.32\columnwidth}
  \centering
  % include first image
\begin{tikzpicture}[]
\begin{axis}[
    ymin=33.5,
    height=3.5cm,
    width=4.5cm,
    axis x line  = bottom,
    axis y line  = left,
    xtick = {0,1,2,3,4,5},
    ytick = {30,35,40,45},
    xticklabels={ 
    $\mathbfcal{E}_3\,\mathbfcal{E}_2\,\mathbfcal{E}_1$,  $\mathbfcal{E}_2\,\mathbfcal{E}_3\,\mathbfcal{E}_1$, $\mathbfcal{E}_3\,\mathbfcal{E}_1\,\mathbfcal{E}_2$, $\mathbfcal{E}_1\,\mathbfcal{E}_3\,\mathbfcal{E}_2$, $\mathbfcal{E}_2\,\mathbfcal{E}_1\,\mathbfcal{E}_3$, $\mathbfcal{E}_1\,\mathbfcal{E}_2\,\mathbfcal{E}_3$},
    ylabel=MRE $\%$,
    xlabel=Sequence of meta environments,
    x label style={at={(0.5,-0.3)}},
    y label style={at={(0.15,0.5)}},
    x tick label style={rotate=45,anchor=east,font=\footnotesize},
     %y tick label style={font=\tiny},
    % axis x line  = bottom,
    % axis y line  = left,
    % ymin=37
    % samples=100,
    % domain=0:10
]

\addplot+[mark=o,color=blue,mark size=1.5pt]           
coordinates                        
{                                   
  (0, 43.08265364202791)  (1, 42.59766434633329)  (2, 41.95367723155698) (3, 41.61097535228704)  (4, 34.41966822592169)  (5, 34.366265862039704)
};

\end{axis}
\end{tikzpicture} 
  %\caption{MRE of meta models for $\mathbfcal{E}_{1}$, $\mathbfcal{E}_{2}$ and $\mathbfcal{E}_{3}$ of \textsc{DeepArch} with size 5.}
  %\caption{\texttt{DeepArch}, S$_{5}$}
  \caption{\textsc{DeepArch}}
  \label{fig:sequence}
\end{subfigure}
~\hspace{0.8cm}
\begin{subfigure}{.32\columnwidth}
  \centering
\begin{tikzpicture}[]
\begin{axis}[
    ymin=10,
    ymax=30,
    height=3.5cm,
    width=4.5cm,
       axis x line  = bottom,
    axis y line  = left,
    xtick = {0,1,2,3,4,5,6,7,8,9,10,11,12,13,14},
    xticklabels={1,2,3,4,5,6,7,8,9},
    ylabel=MRE $\%$,
    xlabel=\# Meta environment,
    x label style={at={(0.5,-0.30)}},
     y label style={at={(0.15,0.5)}},
     %x tick label style={font=\tiny},
     %y tick label style={font=\tiny},
    % axis x line  = bottom,
    % axis y line  = left,
    % ymin=37,
    % samples=100,
    % domain=0:10
]

\addplot+[mark=o,color=blue,mark size=1.5pt]           % 调用绘图函数，并设置绘图的类型是折线图
coordinates                         % 声明是在迪卡尔坐标系中的数据
{                                   % 输入数据
 (0, 27.46) (1, 25.90) (2, 25.31) (3, 23.15) (4, 21.31) (5, 18.10) (6, 16.83) (7, 16.13) (8, 12.67)};

\end{axis}
\end{tikzpicture}
%\caption{MRE of meta models for $\mathbfcal{E}_{10}$ of \textsc{x264} with size 1.}
\caption{\textsc{x264}}
%\caption{\texttt{x264}, $\mathcal{T}_{10}$, S$_{1}$}
% ~\vspace{.03\columnwidth}
  \label{fig:n_meta_tasks}
\end{subfigure}

\vspace{-0.3cm}
\caption{Empirical results that verify the properties of real-world software systems. The y-axis is the testing Mean Relative Error (MRE) on $\mathbfcal{E}_{target}$. (a) confirms Property 1 and 2; $\mathbfcal{E}_{3}$ is the most useful environments for $\mathbfcal{E}_{target}$, following by $\mathbfcal{E}_{2}$ and then $\mathbfcal{E}_{1}$. (b) reveals Property 3.}
 \label{fig:ranking_MRE}
 \vspace{-0.3cm}
\end{figure}
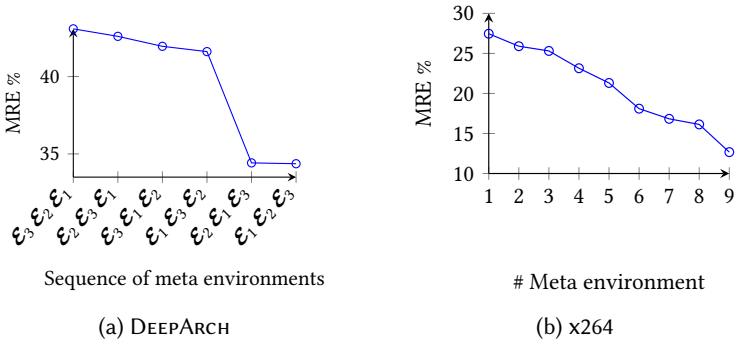

Using the example from Figure~\ref{fig:theory}b again, clearly, the model parameter values learned on the first meta environment $\mathbfcal{E}_1$ will need to be updated via learning the $\mathbfcal{E}_2$ data and $\mathbfcal{E}_3$ data, hence it only contributes to the smallest proportion of its distribution in the meta-model. In contrast, the last meta environment $\mathbfcal{E}_3$ will contribute the most to the meta-model as most of the model parameter values learned under it will be preserved into those of the meta-model (i.e., there will be no further update). $\mathbfcal{E}_2$ contributes less than $\mathbfcal{E}_3$ but more than $\mathbfcal{E}_1$, as its resulted model parameter values are updated once. According to Figure~\ref{fig:theory}a, in this case, $\mathbfcal{E}_3$ has most parts of its parameter distribution close to that of $\mathbfcal{E}_{target}$, hence it is beneficial to leave it as the last to be learned, which preserves more of its model parameter values. Similarly, $\mathbfcal{E}_2$ tends to have more proportions of its parameter distribution closer to that of $\mathbfcal{E}_{target}$ than that of $\mathbfcal{E}_1$, hence learning its data later than that of $\mathbfcal{E}_1$ would ensure it contributes relatively more. All these can then lead to good initialization of the parameter's value in the meta-model (the rightmost in Figure~\ref{fig:theory}b) with respect to the optimal parameter distribution for $\mathbfcal{E}_{target}$ (the rightmost in Figure~\ref{fig:theory}a). Empirically, for a real-world system in Figure~\ref{fig:ranking_MRE}a, placing the more useful meta environments to the latter positions in \Model~can lead to better accuracy for the target environment. Similar results have been observed in other cases.

\texttt{MAML} (Figure~\ref{fig:theory}c), in contrast, forces all meta environments to contribute equally in an aggregation to update the parameter values in the meta-model, hence for configuration data, this can easily result in an initial distribution that is highly deviated from the optimal setting of the target environment (the rightmost in Figure~\ref{fig:theory}c). We will also experimentally justify this in Section~\ref{sec:evaluation}.

\subsection{More Meta Environments are Beneficial}

It is natural to ask, given \textbf{Property 1} and \textbf{Property 2}, why not use only the most useful meta environment to initialize the model parameters for the target environment? The answer is that those less useful meta environments, albeit contributing to relatively fewer proportions in the meta-model, may still provide excellent starting points for certain parts of the parameter distribution. This leads to \Model's final property that fulfills \textbf{Requirement 3}:

\begin{property}
\label{pro:3}
Learning from more meta environments can help to cover the ``corner cases'' of initialing the model parameter values for the target environment. 
%At least, they will not be harmful thanks to \textbf{Property 1} and \textbf{Property 2}.
\end{property}

Considering the example from Figure~\ref{fig:theory}b, we see that, although $\mathbfcal{E}_1$ only preserves a small part of its learned parameter distribution in the meta-model, this may still be close to the optimal setting for the target environment (e.g., the peak parameter values as in Figure~\ref{fig:theory}b), hence complementary to what is missing in the contributions from $\mathbfcal{E}_2$ and $\mathbfcal{E}_3$. The same principle can be similarly applied for $\mathbfcal{E}_2$. As a random example, Figure~\ref{fig:ranking_MRE}b illustrates the sensitivity of \Model's accuracy over a target environment to the number of meta environments (with the appropriate sequence). Clearly, the more meta environments, the better the accuracy---a pattern that we observed for all systems.

Noteworthily, \textbf{Property 3} may not be true for \texttt{MAML} as exploiting all meta environments equally makes it severely suffer from the side-effect of less useful meta environments, which is not uncommon with the configuration data. \Model, in contrast, is able to mitigate such a side-effect by prioritizing the sequence of meta environments, thanks to \textbf{Property 1} and \textbf{Property 2}. Indeed, as we will show in Section~\ref{subsec:rq2}, the sequence-insensitivity has caused \texttt{MAML} to perform significantly worse than \Model~in learning configuration performance data.

%For example, in our experiments  presented in Section~\ref{subsec:rq2} with nine real-world configurable software systems with multiple meta environments, whose side-effects are potentially different, \Model~achieves better average ranks than \texttt{MAML} for all systems.

\section{Implementing and Engineering \Model}
\label{sec:framework}

\begin{figure}[t!]
  \centering
  \includegraphics[width=0.85\columnwidth]{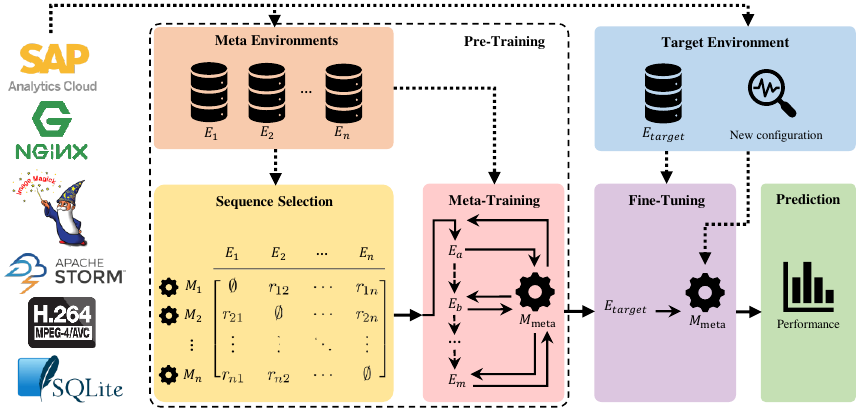}
 \vspace{-0.15cm}
  \caption{The \Model~architecture for learning configuration performance of a system with multiple environments.}
  \label{fig:structure}
   \vspace{-0.5cm}
\end{figure}

%To address the challenges for configuration performance learning under multiple environments and driven by 

To engineer the properties from the theory behind \Model~that fulfill the requirements for configuration performance learning under multiple environments, our implementation has three core components, namely, \textit{Sequence Selection}, \textit{Meta-Training}, and \textit{Fine-Tuning}. The former two resemble a pre-training process for the outer and inner loop (Figure~\ref{fig:models}b) whereas the last is triggered when the target environment becomes available, as shown in Figure~\ref{fig:structure}. These are specified below:

\begin{itemize}
    \item \textbf{Sequence Selection:} This component finds the optimal training sequence of meta environment data in \Model with respect to \textbf{Property 1} and \textbf{Property 2}, considering all the available environments (\textbf{Property 3}). It deals with three low-level questions:

\begin{enumerate}
    \item How to assess the usefulness of the data of a known meta environment with respect to an unforeseen target environment?
    \item How to ensure such an assessment is reliable?
    \item How to guarantee efficiency in the assessment?
\end{enumerate}

     \item \textbf{Meta-Training:} Here, we update the parameters of meta-model by sequentially learning the data of meta environments following the order provided by the \textit{Sequence Selection}.

      \item \textbf{Fine-Tuning:} Given the meta-model from the \textit{Meta-Training}, here the aim is to update the model parameter values using measured configurations under the target environment.
    
\end{itemize}

%While \Model~is agnostic to the base learner, we use \texttt{DeepPerf}~\cite{DBLP:conf/icse/HaZ19}, a regularized Deep Neural Network (rDNN), as the default, since we will show in Section~\ref{sec:evaluation} (and in prior work~\cite{DBLP:conf/icse/HaZ19}) that it achieves the best accuracy amongst the alternatives. However, this base learner can be flexibly replaced when other concern, e.g., training overhead, is of higher priority. 

Given the flexible nature, \Model~is agnostic to the base learner that learns the meta-model, hence it can be paired with any regression learning algorithm from the literature. In the following, we will delineate the above components and the pre-training process of \Model~in Algorithm~\ref{alg:pre-training}.

\subsection{Sequence Selection}

\subsubsection{How to assess the usefulness of meta environments to the target environment?}

Since the target environment is unforeseen by the time of building the meta-model, we assess the overall Mean Relative Error (MRE) on how the single-model $\mathbfcal{M}_i$, which is trained under an individual meta environment $\mathbfcal{E}_i$, performs when being validated across all other remaining meta environments.

Formally, MRE is a widely used scale-free metric for performance prediction~\citep{DBLP:conf/icse/HaZ19, DBLP:conf/esem/ShuS0X20, DBLP:journals/ese/GuoYSASVCWY18}:
\begin{equation}
     MRE = {{1} \over {k}} \times {\sum^k_{t=1} {{|A_t - P_t|} \over {A_t}}} \times 100\%
\end{equation}

    \begin{minipage}{0.54\textwidth}
    \hspace{-0.3cm}
\begin{algorithm}[H]

	\DontPrintSemicolon
	\footnotesize
	%in arbitrary order 
	\caption{Sequence Selection}
	\label{alg:pre-training}
	\KwIn{Data from meta environments $\mathbfcal{E}=\{\mathbfcal{E}_1,\mathbfcal{E}_2,...,\mathbfcal{E}_n\}$}
     \KwOut{The optimal sequence $\mathbfcal{E}_{seq}$}

 %\tcc{\textcolor{blue}{sequence selection.}}
         \For{$\forall \mathbfcal{E}_i \in \mathbfcal{E}$} {

            $\mathcal{M}_i\leftarrow$ \textsc{train($\mathbfcal{E}_i$)}\\

               \For{$\forall \mathbfcal{E}_j \in \mathbfcal{E}$} {

                   \If{$\mathbfcal{E}_j$ is not  $\mathbfcal{E}_i$} {
                         $\mathbf{A}\leftarrow \mathbf{\overline{a}_{ij}}=$ \textsc{testMREwithRepeats($\mathcal{M}_i,\mathbfcal{E}_j$)}\\
                    } \Else{
                      $\mathbf{A}\leftarrow \emptyset$\\
                    }
               }
         }

          \For{$\forall \mathbfcal{E}_j \in \mathbfcal{E}$} {

                %all columns
               \For{$\forall \mathbfcal{E}_i \in \mathbfcal{E}$} {
             
                   $\mathbf{\overline{a}'_j}\leftarrow\mathbf{\overline{a}_{ij}} \in \mathbf{A}$\\
                  
               }
 $\mathbf{\overline{r}_j}=$ \textsc{scottKnottTest($\mathbf{\overline{a}'_j}$)}\\
              \For{$\forall r_{ij} \in \mathbf{\overline{r}_j}: r_{ij} \neq 0$} {
$\mathbf{R}\leftarrow r_{ij}$\\
                }        
         }

 \For{$\forall \text{row } \mathbf{\overline{r}_i} \in \mathbf{R}$} {
 
     $r_{mean}=$ averaging the rank scores $r_{ij} \in \mathbf{\overline{r}_i}$\\
               $\mathbfcal{E}_{seq}\leftarrow \{\mathbfcal{E}_i/\mathcal{M}_i,r_{mean}\}$\\
 }

    \Return $\mathbfcal{E}_{seq}\leftarrow$ \textsc{sort($\mathbfcal{E}_{seq}$)}\\   
	
\end{algorithm}
\end{minipage}
\hspace{-0.2cm}
\begin{minipage}{0.42\textwidth}
\vspace{-2.25cm}
\begin{algorithm}[H]

	\DontPrintSemicolon
	\footnotesize
	
	\caption{Meta-Training}
	\label{alg:meta-training}
	\KwIn{The optimal sequence $\mathbfcal{E}_{seq}$}
     \KwOut{The meta-model $\mathbfcal{M}_{meta}$}

%\tcc{\textcolor{blue}{meta-training.}}
$\mathcal{M}_{meta}\leftarrow$ randomly initialized model \\
 \While{not done}{
         
 %          \If{$\mathcal{M}_{meta} = \emptyset$}{
 % $\mathcal{M}_{meta}\leftarrow$ randomly initialized model \\
         
 %         }

          \For{$\forall \mathbfcal{E}_i \in \mathbfcal{E}_{seq}$} {
                $\mathcal{M}_{meta}\leftarrow$ \textsc{train($\mathcal{M}_{meta}, \mathbfcal{E}_i $)}\\
          }

     }

    \Return $\mathcal{M}_{meta}$\\
	
\end{algorithm}
\begin{algorithm}[H]

	\DontPrintSemicolon
	\footnotesize
	
	\caption{Fine-Tuning}
	\label{alg:fine-tuning}
	\KwIn{The meta-model $\mathbfcal{M}_{meta}$}
     \KwOut{The fine-tuned model $\mathbfcal{M}_{tuned}$}

%\tcc{\textcolor{blue}{meta-training.}}
$\mathbfcal{E}_{target}\leftarrow$ measured data samples from the target environment \\

    \Return   $\mathcal{M}_{tuned}\leftarrow$ \textsc{train($\mathcal{M}_{meta}, \mathbfcal{E}_{target} $)}\\
	
\end{algorithm}
\end{minipage}
\vspace{0.5cm}

\noindent whereby $A_t$ and $P_t$ denote the $t$th actual and predicted performance, respectively. A better overall MRE implies that the distribution of model parameter values learned from $\mathbfcal{E}_i$ 
is generally closer to all the optimal distributions required for fully learning the data of other meta environments, thereby serving as an indicator of the possible usefulness for the unforeseen target environment. As such, $\mathbfcal{E}_i$ will more likely be useful for contributing to the meta-model, enabling quicker adaptation for the unknown target environment.

Algorithm~\ref{alg:pre-training} (lines 2-7) demonstrates the assessment steps below:

%Intuitively, if $\mathbfcal{E}_i$ helps to build a single-model that is a better fit to all the known meta-tasks, then it will be more likely to be useful for contributing to the meta-model, enabling quicker adaptation for the unknown target-task.

%The overall MRE represents how well the distribution of model parameter values learned from $\mathbfcal{E}_i$ can be useful for the other meta-tasks in general, as
\begin{enumerate}
    \item Train a single-model $\mathbfcal{M}_i$ using all the available data for a meta environment $\mathbfcal{E}_i$ (line 2).

    \item Test single-model $\mathbfcal{M}_i$ over all data for the remaining meta environments $\mathbfcal{E}_j$. This is repeated $x$ times ($x=30$ in this work as we found that using more repeats did not change the result), leading to $x$ MRE values for each tested meta environment, denoted as a vector $\mathbf{\overline{a}_{ij}}$ (line 3-7).
    \item All the $\mathbf{\overline{a}_{ij}}$ are represented in a matrix $\mathbf{A}$ below:
%\[
\begin{equation}
\mathbf{A} = 
\begin{blockarray}{c cccc}
& \mathbfcal{E}_1  & \mathbfcal{E}_2 & \cdots & \mathbfcal{E}_n  \\
%\cmidrule{2-6} \cmidrule{8-12}
\cmidrule{2-5}
\begin{block}{c [cccc]}
\mathbfcal{M}_1 & \fixhd{b} \emptyset & \mathbf{\overline{a}_{12}} & \cdots & \mathbf{\overline{a}_{1n}}  \\
\mathbfcal{M}_2 & \fixhd{b} \mathbf{\overline{a}_{21}} & \emptyset & \cdots & \mathbf{\overline{a}_{2n}}  \\
\vdots & \vdots & \vdots & \ddots & \vdots  \\
\mathbfcal{M}_n & \fixhd{b} \mathbf{\overline{a}_{n1}} & \mathbf{\overline{a}_{n2}} & \cdots & \emptyset \\
\end{block}
\noalign{\vspace{-100.5ex}}
\\
\end{blockarray}
\end{equation}
%\]

 \item Repeat from (1) till there is a single-model for every meta environment.
    %\item Repeat from (1) for the next meta-task until there is a single-model for all.
\end{enumerate}

\subsubsection{How to ensure the reliability of assessment?}

A naive way to sort the sequence would be to directly use the overall MRE as the metric in the comparison. This, however, entails two issues:

\begin{itemize}

\item Since the overall MRE covers the accuracy tested over all the remaining meta environments, the comparisons may fail to consider statistical significance even with repeated runs.

%\item The MRE tested on each individual meta-task can be largely deviated, e.g., we observed that some meta-tasks result in a MRE between $1\%$ and $20\%$ while some others range from $200\%$ to $500\%$. Standardization is possible but it will inevitably lose the characteristics of distributions.

\item Due to the residual nature of MRE, the single-model $\mathbfcal{M}_i$ has to be built by the base learner, which may not be realistic when the training is expensive (see Section~\ref{sec:efficiency}).

\end{itemize}
%\footnote{All the sets of MRE with $\emptyset$ are ignored and given a rank of 0.}

Instead, in \Model~we use Scott-Knott test~\citep{DBLP:journals/tse/MittasA13} to rank the MREs of all single-models tested on one meta environment, then average the ranks for a single-model $\mathbfcal{M}_i$ when testing it on all the remaining meta environments. Scott-Knott sorts the list of treatments (e.g., $\mathbf{\overline{a}_{1j}},\mathbf{\overline{a}_{2j}},...,\mathbf{\overline{a}_{nj}}$ from all single-models tested on the $j$th meta environment) by their average MRE. Next, it splits the list into two sub-lists with the largest difference $\Delta$ ~\cite{xia2018hyperparameter}:
\begin{equation}
    \Delta = \frac{|l_1|}{|l|}(\overline{l_1} - \overline{l})^2 + \frac{|l_2|}{|l|}(\overline{l_2} - \overline{l})^2
\end{equation}
whereby $|l_1|$ and $|l_2|$ are the sizes of two sub-lists ($l_1$ and $l_2$) from list $l$ with a size $|l|$. $\overline{l_1}$, $\overline{l_2}$, and $\overline{l}$ denote their mean MRE. During the splitting, bootstrapping and $\hat{A}_{12}$~\citep{Vargha2000ACA} are applied to check if $l_1$ and $l_2$ are significantly different. If that is the case, Scott-Knott recurses on the splits. In other words, the models are divided into different sub-lists if both bootstrap sampling suggests that a split is significant under a confidence level of 99\% and with a good effect $\hat{A}_{12} \geq 0.6$. The sub-lists are then ranked based on their mean MRE. For example, when comparing $A$, $B$, and $C$, a possible split could be $\{A, B\}$, $\{C\}$, with the rank score of 1 and 2, respectively. Hence, statistically, we say that $A$ and $B$ perform similarly, but they are significantly better than $C$.

%whereby $l_1$ and $l_2$ are the sub-list of list $l$;  $|l_1|$, $|l_2|$, and  $|l|$ are their size and $\overline{l_1}$, $\overline{l_2}$, and $\overline{l}$ denote their mean MRE.

%Formally, Scott-Knott test finds the best split by maximizing the difference $\Delta$ in the expected mean before and after each split:

%In other words, the approaches are divided into different sub-lists if both bootstrap sampling and effect size test suggest that a split is statistically significant (with a confidence level of 99\%) and with a good effect $\hat{A}_{12} \geq 0.6$. The sub-lists are then ranked based on their mean MRE.

Specifically, we compute the final rank scores as in Algorithm~\ref{alg:pre-training} (lines 9-17):

\begin{enumerate}
    \item Get the repeated MREs for all single-models tested for a meta environment $\mathbfcal{E}_j$ (lines 9-10).
    \item Compute the vector of rank scores $\mathbf{\overline{r}_j}$  for all single-models when testing on $\mathbfcal{E}_j$ (line 11).
    \item Replace the tested result for $\mathbfcal{E}_j$ in matrix $\mathbf{A}$ from a vector $\mathbf{\overline{a}_{ij}}$ to a single value of rank score $r_{ij}$, forming a new matrix of ranks $\mathbf{R}$. (lines 12-13).
    \item Repeat from (1) until all tested meta environments are covered.
    \item Compute the average rank ($\mathbf{\overline{r}_i}$) over all tested meta environments for each single-model $\mathbfcal{M}_i$, and put it with the corresponding meta environment learned by $\mathbfcal{M}_i$ in $\mathbfcal{E}_{seq}$ (lines 14-16).
\end{enumerate}

The optimal sequence of training the meta environments can be found by sorting the average rank scores in $\mathbfcal{E}_{seq}$ descendingly (line 17)---the most useful meta environment is left to the last.

\subsubsection{How to guarantee efficiency?}
\label{sec:efficiency}

Ideally, the single-model $\mathbfcal{M}_i$ used when assessing the usefulness of the meta environments for a target environment should be also learned by the base learner. Yet, training some base learners, such as DNN, can be expensive, especially when we need to train a single-model for each meta environment. To ensure efficiency, in \Model, we use linear regression as a surrogate of the base learner to build $\mathbfcal{M}_i$ since it has negligible overhead. 

Indeed, the MRE of linear regression can differ from that of the base learner. However, through Scott-Knott test, we are mainly interested in the relative ranks between the MREs from different single-models rather than their residual accuracy. As a result, simple models like linear regression can still produce similarly coarse-grained ranks to that of a complex model~\cite{DBLP:conf/sigsoft/NairMSA17}.

Although the sequence selection needs to be rerun for every new target environment, linear regression renders the process very fast, ranging from 2 to 60 seconds for 2 to 9 meta environments.
% The extensibility of the method could be more discussed.

\subsection{Meta-Training}

%Once the optimal sequence of meta environments has been determined, 

The \textit{Meta-Training} in \Model~learns the dataset of each meta environment sequentially according to the optimal sequence (Algorithm~\ref{alg:meta-training}). Notably, the model parameter values trained under a preceding meta environment serve as the starting point for learning the data of the succeeding one. 

In this work, we use \texttt{DeepPerf}~\cite{DBLP:conf/icse/HaZ19}---a regularized Deep Neural Network (rDNN)---as the default base learner since it is a state-of-the-art model used for configuration performance learning. However, we would like to stress that \Model~is agnostic to the base learner, hence the choice of using \texttt{DeepPerf} is mainly due to its empirical superiority on accuracy over the others~\cite{DBLP:conf/icse/HaZ19}. The base learner can be replaced when other concerns, e.g., training overhead, are of higher priority.

%Further, as we will show in Section~\ref{sec:evaluation} (and as in prior work~\cite{DBLP:conf/icse/HaZ19}), it tends to outperform the other peer models due to its ability in handling feature sparsity within the configuration data.

%While \Model~is agnostic to the base learner, we use \texttt{DeepPerf}~\cite{DBLP:conf/icse/HaZ19}, a regularized Deep Neural Network (rDNN), as the default, since we will show in Section~\ref{sec:evaluation} (and in prior work~\cite{DBLP:conf/icse/HaZ19}) that it achieves the best accuracy amongst the alternatives. However, this base learner can be flexibly replaced when other concern, e.g., training overhead, is of higher priority.

\texttt{DeepPerf} is trained in the same process as used by~\citet{DBLP:conf/icse/HaZ19} with hyperparameter tuning for building the single-models. We kindly refer interested readers to their work for details.

%including the number of neurons, the learning rate, and parameters for the $L_1$ regularization.

\subsection{Fine-Tuning}

Since we seek to exploit the learned model parameter values from the trained meta-model as the starting point in fine-tuning, \Model~adopts the same \texttt{DeepPerf} as the base learner for the target environment. As in Algorithm~\ref{alg:fine-tuning}, the fine-tuning follows standard training of a typical machine learning model using samples for the target environment; yet, instead of using an initial model with random parameter values, the meta-model directly serves as the starting point. As a result, learning under the target environment can be greatly expedited and improved as the knowledge from meta environments should have been generalized. The training sample size from the target environment can vary, for which we examine different sizes as what will be discussed in Section~\ref{sec:setup}. The newly given configurations can be fed into the fine-tuned meta-model to predict their performance.
%As with the meta-training, the same hyperparameter tuning procedure is used during the process.

%\section{Experiment Setup}
\section{Experiment Setup}
\label{sec:setup}

%This section delineates the settings of our experimental evaluation. 

%\subsection{Research Questions}

Our experiments evaluate~\Model~by answering the following research questions (RQs):

\begin{itemize}
    \item \textbf{RQ$_1$:} How does \Model~perform compared with existing single environment models?

    \item \textbf{RQ$_2$:} How does \Model~perform compared with the state-of-the-art models that handle multiple environments?

    \item \textbf{RQ$_3$:} How effective is the sequence selection in \Model?

    \item \textbf{RQ$_4$:} What is the sensitivity of \Model's accuracy to the sample size used in pre-training?
\end{itemize}

%We ask RQ$_1$ to examine whether the idea of ``learning-to-learn'' achieved by meta-learning is more effective than learning only on the target tasks as in existing single-environment models. We answer RQ$_2$ to check if the concept of sequential learning is effective for learning software performance compared with existing transfer learning, multi-task learning, and meta-learning models, including those general ones and those that are specific to performance learning. Through different base learners, we study RQ$_3$ to validate whether it is necessary to have our specialized design of the pre-training in \Model, instead of merely providing the data of all meta-tasks in random order. Finally, we seek to understand how the training sample size of the meta-tasks in the pre-training phase affects the results of \Model~via asking RQ$_4$.
\begin{table}[t!]
\caption{Details of the subject systems and the training sizes (for fine-tuning) when an environment is used as the target. $|\mathbfcal{H}|$, $|\mathbfcal{W}|$, and $|\mathbfcal{V}|$ respectively denotes the number of hardware, workload, and version considered for the environments, ($|\mathbfcal{B}|$/$|\mathbfcal{N}|$) denotes the number of binary/numerical options, and $|\mathbfcal{C}|$ denotes the number of valid configurations per environment (full sample size).}
\vspace{-0.3cm}
\centering
\footnotesize
\begin{adjustbox}{width=\linewidth,center}
\setlength{\tabcolsep}{1mm}
\begin{tabular}{l|llll|ccc|lllll}
\toprule

\multirow{2}{*}{\textbf{System}} & \multirow{2}{*}{\textbf{Domain}}  & \multirow{2}{*}{\textbf{$|\mathbfcal{B}|$/$|\mathbfcal{N}|$}} & \multirow{2}{*}{\textbf{$|\mathbfcal{C}|$}}  & \multirow{2}{*}{\textbf{Used by}}    & \multicolumn{3}{c|}{\textbf{Environments}}    & \multicolumn{5}{c}{\textbf{Training Size}}  \\

&   &  & &   & \textbf{$|\mathbfcal{H}|$} & \textbf{$|\mathbfcal{W}|$} & \textbf{$|\mathbfcal{V}|$} & \textbf{$S_1$} & \textbf{$S_2$} & \textbf{$S_3$} & \textbf{$S_4$} & \textbf{$S_5$}    \\
\midrule
\textsc{DeepArch} & DNN tool            & 12/0             & 4096  & \cite{DBLP:conf/sigsoft/JamshidiVKS18}& 3 & 1 & 1  & 12  & 24 & 36 & 48 & 60\\
\textsc{SaC}      & Cloud tool          & 58/0             & 4999  & \cite{DBLP:journals/corr/abs-1911-01817}       & 1 & 3 & 1     & 58 & 116 & 174 & 232 & 290\\
\textsc{SQLite}   & DBMS                & 14/0             & 1000  & \cite{DBLP:journals/corr/abs-1911-01817}     & 2  & 1 & 2       & 14 & 28 & 42 & 56 & 70\\
% \textsc{GCC}      & Compilation    & 5       & 5/0              & 81    & \cite{DBLP:journals/corr/abs-2112-07279}     \\
\textsc{NGINX}      & Web server          & 16/0              & 1104    & \cite{DBLP:conf/icse/webertwins}   & 1 & 1 & 4   & 16 & 32 & 48 & 64 & 80\\
\textsc{SPEAR}    & Audio editor        & 14/0             & 16385 & \cite{DBLP:journals/corr/abs-1911-01817}   & 3 & 1 & 1       & 14 & 28 & 42 & 56 & 70\\
\textsc{Storm} & Big data analyzer         & 1/11             & 2048  & \cite{DBLP:conf/sigsoft/JamshidiVKS18} & 3 & 1 & 1  & 158 & 211 & 522 & 678 & 1403\\
\textsc{ImageMagick}     & Image editor      & 0/5            & 100   & \cite{LESOIL2023111671} & 1 & 4 & 1       & 11 & 24 & 45 & 66 & 70 \\
% \textsc{XZ}     & Data compression & 4      & 0/4            & 30   & \cite{DBLP:journals/corr/abs-2112-07279}      \\
\textsc{ExaStencils}    & Code generator      & 8/4            & 4098   & \cite{DBLP:conf/icse/webertwins} & 1  & 4 & 1      & 106 & 181 & 366 & 485 & 695\\ 
\textsc{x264}     & Video encoder      & 11/13            & 201   & \cite{LESOIL2023111671}    & 1 & 10 & 1   & 24 & 53 & 81 & 122 & 141\\

\bottomrule
\end{tabular}
\end{adjustbox}
\label{chap-meta-table:subject systems}
\vspace{-0.3cm}
\end{table}

\subsection{Subject Software Systems and Environments}
\label{subsec:subject_system}

In the experiments, we study widely-used multi-environment configuration datasets collected from real-world systems\footnote{To ensure reliability, the data has been collected with repeated measurements.}~\cite{DBLP:conf/kbse/JamshidiSVKPA17, DBLP:conf/sigsoft/JamshidiVKS18, DBLP:journals/corr/abs-1911-01817, LESOIL2023111671}. These systems are selected based on the following:

\begin{itemize}
    \item To ensure diversity, systems with less than three environments are removed.
    \item In the presence of multiple datasets for the same system, the one that contains the most deviated measurements between the environments is used.
\end{itemize}

Within the identified systems, we rule out the data of environments that (1) do not measure all valid configurations; (2) contain invalid measurements; or (3) lack detailed specifications.

% \begin{itemize}
%     \item Not all valid configurations are measured.
%     \item There are invalid measurements.
%     \item The specification of the environment is missing.
% \end{itemize}

%All systems come from diverse domains and can either be binary or a mix of both binary and categorical/numerical options. To reduce noise, systems that contain invalid configurations or with less than three environments are removed. In particular, the environments for each system are chosen according to the following criteria:

%Table~\ref{chap-meta-table:subject systems} presents the details of the configurable software systems studied, which come from different domains.

As shown in Table~\ref{chap-meta-table:subject systems}, the above process leads to nine systems of diverse domains, scales, and option types, together with configurations measured in distinct valid environments of hardware, workloads, and/or versions. To further improve the external validity, we examine the models under different training sample sizes from the target environment. For binary systems (only binary options), we use five sizes as prior work~\cite{DBLP:conf/icse/HaZ19}: $\{x,2x,3x,4x,5x\}$ where $x$ is the number of options. For mixed systems (both binary and numeric options), we use five sizes as produced by the sampling method from the work of~\citet{DBLP:journals/sqj/SiegmundRKKAS12}. These are the common methods used in the field~\cite{DBLP:conf/icse/HaZ19,DBLP:conf/esem/ShuS0X20}. 

%To ensure generalizability, all systems come from diverse domains and can either be binary, i.e., with only binary options (e.g., \textsc{SaC}) or mixed that contains both binary and categorical/numerical options, e.g., \textsc{x264}, which is more difficult to model~\citep{DBLP:conf/icse/HaZ19}.

%Each of the systems contains data collected under environments with distinct running conditions, including hardware, version, and workload. 

%Different from the single-environment performance learning tasks, the multi-environment tasks are collected under different running conditions, for example, the environments for \textsc{SQLite} include 3 combinations of 2 hardware conditions (Azure and AWS cloud platform), and 2 software versions (1.2 and 2.7)~\citep{DBLP:conf/kbse/JamshidiSVKPA17, DBLP:journals/corr/abs-1911-01817}, and the environments for \textsc{x264} are from 10 distinct workloads~\citep{DBLP:journals/corr/abs-2112-07279}.

%The results are in Table~\ref{tb:sizes}.

%The training sample sizes used are shown in Table~\ref{tb:sizes}, where for the binary systems the sizes are from 1 to 5 times the number of features~\cite{DBLP:conf/icse/HaZ19}, while for the mix systems, the sizes are decided by the sampling method used by~\cite{DBLP:journals/sqj/SiegmundRKKAS12}. These training sizes are chosen because they have been widely used in performance prediction studies. 

%\input{Tables/sample_sizes.tex}

% \input{Tables/chap-DAL/sample_sizes.tex}

\subsection{Procedure, Metric and Statistical Validation}

% The pre-training, however, is run only once with a single iteration of the outer loop---the same setting as in existing meta-leanring work [].

\subsubsection{Procedure}

For each system and comparative approach, we follow the steps below:

\begin{enumerate}
    \item Pick an environment as the target $\mathbfcal{E}_{target}$ and set all remaining ones as meta environments.
    %(because of \textbf{Property 3}).
    \item If applicable, pre-training on all data samples of the meta environments (except for RQ$_4$).
    \item Pick a training sample size $S_i$ for $\mathbfcal{E}_{target}$.
    %\item Fine-tune for $\mathbfcal{E}_{target}$ under $S_i$;
    \item Train/fine-tune a model under $S_i$ randomly selected samples and test it on all the remaining (unforeseen) samples of $\mathbfcal{E}_{target}$. 
    \item To mitigate bias, repeat (4) for 30 runs via bootstrapping without replacement\footnote{30 runs is the most common setting in the field of configuration performance learning~\cite{DBLP:conf/icse/HaZ19,DBLP:conf/esem/ShuS0X20}. This number of runs is merely a pragmatic choice given the resource constraint.}.
    \item Repeat from (3) to cover all training sample sizes of $\mathbfcal{E}_{target}$.
    \item Repeat from (1) till every environment has served as the target environment once. In this way, we avoid bias towards the prediction of a particular environment.
\end{enumerate}

As such, for each of the nine systems, we have five different training sizes and 3--10 alternative target environments, leading to 15--50 cases of comparisons. In \Model, we set the iteration limit for the outer loop as 1, which is the same default for \texttt{MAML}. 

Noteworthily, in this work, we assume the configurations are the same for different meta environments for all approaches, which is derived from the existing datasets and is the common assumption from existing work, e.g., BEETLE~\cite{DBLP:journals/corr/abs-1911-01817}. However, since the learning of data from the meta environments is independent of each other in \Model~(i.e., the information sharing happens at the model parameter level rather than at the data level), it is not restricted to this assumption and can directly learn data from the meta-environments for which the configurations in the data samples are not the same.

%use each task as the target-task once and the reaming serve as the meta-tasks/sources. All the available data samples of the meta-tasks are used in the pre-training, unless otherwise stated. To mitigate bias, all experiments (i.e., the fine-tuning or single-task training) are repeated for 30 runs via bootstrapping without replacement. 

\subsubsection{Accuracy}

As with the sequence selection, we use MRE to assess the accuracy, which is the standard practice~\cite{DBLP:conf/splc/ValovGC15,DBLP:conf/oopsla/QueirozBC16,DBLP:conf/icse/0003XC021,DBLP:conf/icse/HaZ19,DBLP:journals/ese/GuoYSASVCWY18}. Further, MRE is insensitive to the scale of performance metrics.

\subsubsection{Speedup}

As in prior work~\cite{DBLP:conf/icse/0003XC021,DBLP:conf/sigsoft/0001L21,DBLP:conf/sigsoft/0001L24}, to assess the training speedup achieved by \Model~for a system, we use a baseline,
$b$, taken as the smallest training size that the best
state-of-the-art counterpart consumes to achieve its best mean MRE. We then find the smallest size for \Model~to achieve the same accuracy, denoted as $s$. The ratios, i.e., $sp={b \over s}$, are reported. Clearly, if \Model~is more data-efficient, then we would expect $sp \geq 1 \times$. When \Model~cannot achieve the same mean MRE for all sample sizes, we have $sp=N/A$.

\subsubsection{Statistics}

To make a fair comparison between more than two models, we again use the Scott-Knott test~\citep{DBLP:journals/tse/MittasA13} to evaluate their statistical significance on the difference of MRE over 30 runs, as recommended by~\citet{DBLP:journals/tse/MittasA13}. In a nutshell, Scott-Knott test provides the following benefits over the other statistical methods:

\begin{itemize}
    \item Unlike those parametric tests such as the t-test~\cite{student1908probable}, it has no assumption on the data distribution.
    \item It allows the assessment of more than two models as opposed to the other pair-wise nonparametric tests such as the Wilcoxon test~\cite{Wilcoxon1945IndividualCB} or Mann-Whitney U test~\cite{Mann47}.
    \item It naturally ranks the model while other multiple comparison tests, e.g., Kruskal-Wallis test~\cite{mckight2010kruskal}, only assess the differences between models without comprising the better or worse. Further, the Scott-Knott test does not need post-hoc correlation.
\end{itemize}

\section{Evaluation}
\label{sec:evaluation}

We now present and discuss the experiment results for the RQs.

\subsection{\Model~against Single Environment Models}
\subsubsection{Method}

To answer \textbf{RQ}$_1$, we compare \Model~against five state-of-the-art single environment models, i.e., \texttt{DeepPerf}, \texttt{DECART}, \texttt{RF}, \texttt{SPLConqueror}, and \texttt{XGBoost}, as discussed in Section~\ref{sec:bg-single}. To ensure fairness, we use the code published by their authors. All settings from Section~\ref{sec:setup} are used.

% \begin{itemize}
%     \item \textbf{\texttt{DeepPerf}}~\citep{DBLP:conf/icse/HaZ19}: a  regularized DNN model with fast hyperparameter tuning for performance learning. This is also used as the default base learner in \Model. 
%     %It has been compared as a SOTA performance model in most of the recent studies since 2019~\citep{DBLP:conf/esem/ShuS0X20}.
%     \item \textbf{\texttt{DECART}}~\citep{DBLP:journals/ese/GuoYSASVCWY18}: an improved CART model with speciflized sampling machnism.
%     \item \textbf{\texttt{Random forest} (\texttt{RF})}~\citep{DBLP:conf/splc/ValovGC15,DBLP:conf/oopsla/QueirozBC16,DBLP:conf/icse/0003XC021}: a commonly used in approach to tackle the feature sparsity issue with good resilience to overfitting.
%     \item \textbf{\texttt{SPLConqueror}}~\citep{DBLP:journals/sqj/SiegmundRKKAS12}: a baseline approach that relies on linear regression.
% \end{itemize}

\subsubsection{Results}

% \begin{table*}[]
% % \begin{sidewaystable}[]
% \footnotesize
% % \setlength{\tabcolsep}{1mm}
% \renewcommand\arraystretch{1.2}
% \caption{The average rank, the mean and interquartile range of MRE of the 5 sizes, denoted as Mean (IQR), for \Model~and the single-task approaches for all the subject systems and training sizes over 30 runs. For each case, the one(s) with the best rank from the Scott-Knott test is highlighted in bold, and the \setlength{\fboxsep}{1.5pt}\colorbox{green!20}{\textbf{green cells}} mean \Model~is ranked the first.}
% % \begin{adjustbox}{width=\textwidth,center}
% \input{Tables/RQ1.tex}
% % \end{adjustbox}
% \label{tb:RQ1}
% % \vspace{-0.4cm}
% % \end{sidewaystable}
% \end{table*}

%\textcolor{red}{up to 81\% (\textsc{Exastencils}, $S_{1}$), 84\% (\textsc{SaC}, $S_{5}$), 88\% (\textsc{NGINX}, $S_{2}$), and 99\% (\textsc{NGINX}, $S_{5}$)}

%In some systems (\textsc{SPEAR} and \textsc{Storm})

  \begin{figure}[!t]
  \centering
  \begin{subfigure}[t]{0.65\columnwidth}
        \centering
\includegraphics[width=\textwidth]{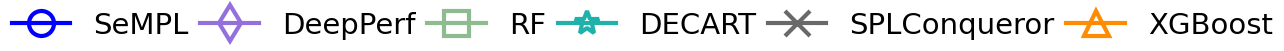}
   \end{subfigure}
%% Legend %%

%    \begin{subfigure}[t]{\subfigsize\columnwidth}  
%    \vspace{-0.6cm}
%         \centering
%  \subcaption*{\scriptsize$sp=2.25\times$}
%    \end{subfigure}
% ~\hspace{-0.3cm}
% \begin{subfigure}[t]{\subfigsize\columnwidth}
% \vspace{-0.6cm}
%         \centering
%    \subcaption*{\scriptsize$sp=1.35\times$}
%    \end{subfigure}
% ~\hspace{-0.3cm}
%    \begin{subfigure}[t]{\subfigsize\columnwidth}
%    \vspace{-0.6cm}
%         \centering
%     \subcaption*{\scriptsize$sp=1.58\times$}
%    \end{subfigure}

   \begin{subfigure}[t]{\subfigsize\columnwidth}
        \centering
\includegraphics[width=\textwidth]{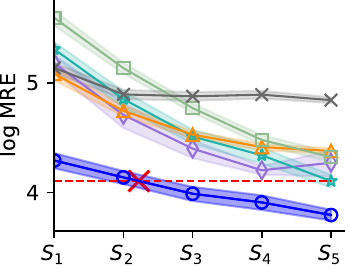}
 \subcaption*{\textsc{DeepArch}  ($sp=2.25\times$)}
   \end{subfigure}
~\hspace{-0.15cm}
\begin{subfigure}[t]{\subfigsize\columnwidth}
        \centering
\includegraphics[width=\textwidth]{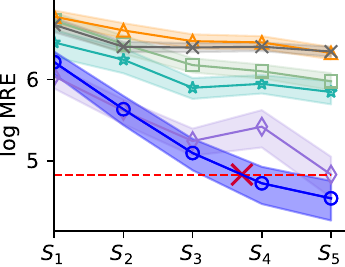}
      \subcaption*{\textsc{SaC}  ($sp=1.35\times$)}
   \end{subfigure}
~\hspace{-0.15cm}
   \begin{subfigure}[t]{\subfigsize\columnwidth}
        \centering
\includegraphics[width=\textwidth]{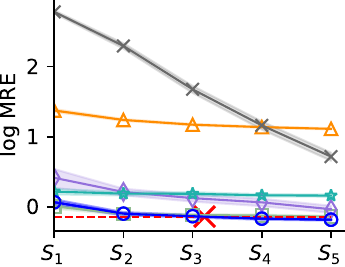}
     \subcaption*{\textsc{SQLite}  ($sp=1.58\times$)}
   \end{subfigure}
~\hspace{-0.15cm}
 \begin{subfigure}[t]{\subfigsize\columnwidth}
        \centering
\includegraphics[width=\textwidth]{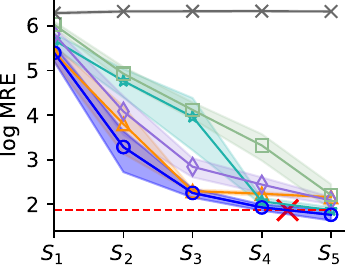}
 \subcaption*{\textsc{NGINX} ($sp=1.14\times$)}
   \end{subfigure}

\begin{subfigure}[t]{\subfigsize\columnwidth}
        \centering
\includegraphics[width=\textwidth]{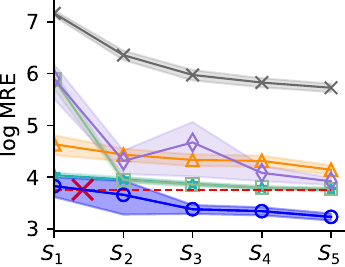}
     \subcaption*{\textsc{SPEAR} ($sp=3.54\times$)}
   \end{subfigure}
~\hspace{-0.15cm}
    \begin{subfigure}[t]{\subfigsize\columnwidth}
        \centering
\includegraphics[width=\textwidth]{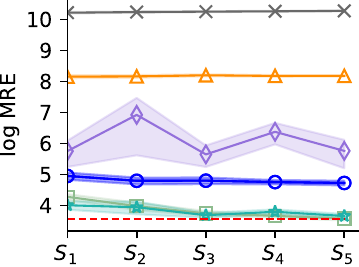}
      \subcaption*{\textsc{Storm}  ($sp=N/A$)}
   \end{subfigure}
~\hspace{-0.15cm}
      \begin{subfigure}[t]{\subfigsize\columnwidth}
        \centering
\includegraphics[width=\textwidth]{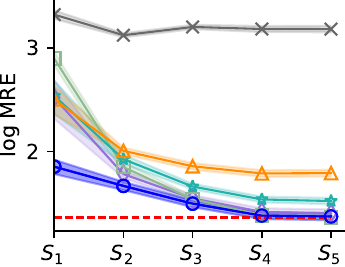}
 \subcaption*{\textsc{ImageMagick}  ($sp=N/A$)}
   \end{subfigure}
~\hspace{-0.15cm}
   \begin{subfigure}[t]{\subfigsize\columnwidth}
        \centering
\includegraphics[width=\textwidth]{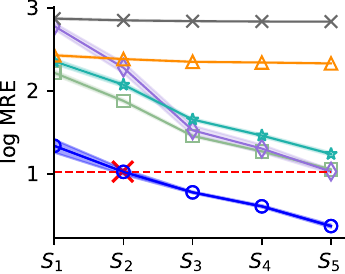}
     \subcaption*{\textsc{ExaStencils}  ($sp=3.86\times$)}
   \end{subfigure}

    \begin{subfigure}[t]{\subfigsize\columnwidth}
        \centering
\includegraphics[width=\textwidth]{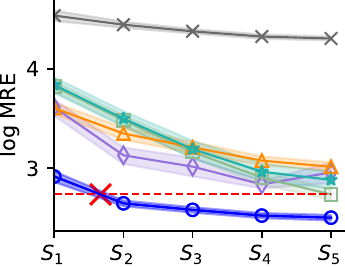}
      \subcaption*{\textsc{x264}  ($sp=3.25\times$)}
   \end{subfigure}
~\hfill
 \begin{subfigure}[t]{0.7\columnwidth}
 \vspace{-2.2cm}
  %\begin{table}[!t]
%\caption{Average Scott-Knott ranks for \Model~versus multi-environment models over all cases/runs. \setlength{\fboxsep}{1.5pt}\colorbox{steel!20}{blue} is the best.}
\centering
\footnotesize
\begin{adjustbox}{width=\linewidth,center}
\begin{tabular}{l|ccccccccc}
\toprule
\textbf{Model} & \rotatebox[origin=c]{0}{\textsc{DeepArch}} & \rotatebox[origin=c]{0}{\textsc{SaC}} & \rotatebox[origin=c]{0}{\textsc{SQLite}} & \rotatebox[origin=c]{0}{\textsc{NGINX}} & \rotatebox[origin=c]{0}{\textsc{SPEAR}} & \rotatebox[origin=c]{0}{\textsc{Storm}} & \rotatebox[origin=c]{0}{\textsc{ImageMagick}} & \rotatebox[origin=c]{0}{\textsc{ExaStencils}} & \rotatebox[origin=c]{0}{\textsc{x264}}\\ 
\midrule
\Model  & \cellcolor{steel!20}\textbf{1.0}  & \cellcolor{steel!20}\textbf{1.53}  & \cellcolor{steel!20}\textbf{1.15}  & \cellcolor{steel!20}\textbf{1.45}  & \cellcolor{steel!20}\textbf{1.07}  & 3.2  & \cellcolor{steel!20}\textbf{1.3}  & \cellcolor{steel!20}\textbf{1.0}  & \cellcolor{steel!20}\textbf{1.02}  \\ 
\texttt{DeepPerf}  & 2.4  & 2.27  & 2.8  & 2.55  & 3.13  & 3.53  & 1.9  & 3.2  & 2.26  \\ 
\texttt{RF}  & 4.0  & 2.47  & 1.4  & 3.45  & 2.13  & \cellcolor{steel!20}\textbf{1.13}  & 2.05  & 2.1  & 2.78  \\ 
\texttt{DECART}  & 2.87  & 1.67  & 3.15  & 2.4  & 1.87  & 1.73  & 3.05  & 3.35  & 3.18  \\ 
\texttt{SPLConqueror}  & 4.0  & 4.2  & 5.05  & 4.65  & 4.4  & 5.87  & 5.0  & 5.7  & 4.62  \\ 
\texttt{XGBoost}  & 2.93  & 3.87  & 4.6  & 1.95  & 3.0  & 4.87  & 3.8  & 4.5  & 3.04  \\ 

% \Model  & \cellcolor{steel!20}\textbf{1.0}  & \cellcolor{steel!20}\textbf{1.53}  & \cellcolor{steel!20}\textbf{1.15}  & \cellcolor{steel!20}\textbf{1.2}  & \cellcolor{steel!20}\textbf{1.07}  & 3.2  & \cellcolor{steel!20}\textbf{1.3}  & \cellcolor{steel!20}\textbf{1.0}  & \cellcolor{steel!20}\textbf{1.02}  \\ 
% \texttt{DeepPerf}  & 2.33  & 2.27  & 2.8  & 2.25  & 2.87  & 3.53  & 1.9  & 3.05  & 2.2  \\ 
% \texttt{RF}  & 3.8  & 2.47  & 1.4  & 3.1  & 2.0  & \cellcolor{steel!20}\textbf{1.13}  & 2.1  & 2.1  & 2.54  \\ 
% \texttt{DECART}  & 2.73  & 1.67  & 3.15  & 2.15  & 1.87  & 1.73  & 3.1  & 3.35  & 2.9  \\ 
% \texttt{SPLConqueror}  & 3.73  & 3.6  & 4.4  & 4.2  & 3.87  & 4.87  & 4.3  & 4.75  & 4.04  \\ 

\bottomrule
\end{tabular}
\end{adjustbox}
%\label{tb:rank_rq3}
%\end{table}
   \vspace{0.14cm}
\subcaption*{Mean Scott-Knott ranks over all cases/runs. \setlength{\fboxsep}{1.5pt}\colorbox{steel!20}{blue} cell denotes the best.}
 \label{fig:mappings}
 % \vspace{-0.4cm}
    \end{subfigure}
%~\hspace{1cm}
     
    %\caption{The log mean MRE of the compared performance models for 5 training size and 9 subject systems in RQ$_{1}$.}
      \caption{\Model~versus single environment models. For the simplicity of exposition, we report the log-transformed average MRE (and its standard error) of all target environments and runs. For speedup ($sp={b \over s}$), \textcolor{red}{\dashed} denotes the mean MRE for $b$; \textcolor{red}{\ding{53}} indicates the point of $s$. Detailed data can be accessed at:  \texttt{\textcolor{blue}{\protect\url{https://github.com/ideas-labo/SeMPL/blob/main/Figure5_full.pdf}}}.}
      \label{fig:rq1-mean}
      \vspace{-0.3cm}
  \end{figure}

As can be seen in Figure~\ref{fig:rq1-mean}, on nearly all systems, \Model~significantly improves the MRE by up to 81\%, 87\%, 84\%, 99\% and 91\% over \texttt{DeepPerf}, \texttt{RF}, \texttt{DECART}, \texttt{SPLConqueror} and \texttt{XGBoost}, respectively. As the default base learner for \Model, \texttt{DeepPerf} might occasionally perform worse even with more data. This is because the target environments exhibit diverse data patterns, which can easily cause the hyperparameter tuning to be trapped at local optima. Such an issue can be mitigated when paired with \Model~since the meta-model is fine-tuned from some good starting points of the parameters. In particular, \Model~often achieves considerably better MRE in very few data samples, commonly with a significant speedup that can be up to $3.86\times$. This confirms the generalization efficiency of sequential meta-learning. Remarkably, from the average Scott-Knott ranks, \Model~is the best for 8 out of 9 systems (i.e., $89\%$), demonstrating its superiority over the single environment models for learning configuration performance, thanks to ``learning to learn".

Noteworthy, \Model~does not achieve the highest rank for \textsc{Storm}, which can be attributed to a combination of factors including the sparsity of samples (due to the nature of the system) and the presence of more diverse environments, primarily limited to hardware changes.

\begin{quotebox}
   \noindent
   \textit{\textbf{RQ$_{1}$:} \Model~performs significantly better than the state-of-the-art single environment performance models in 8 out of 9 systems with the best MRE improvement of $99\%$; it is also generally data-efficient with up to $3.86\times$ speedup.} 
\end{quotebox}

\subsection{\Model~against Multi-Environment Models}
\label{subsec:rq2}

\subsubsection{Method}

To study \textbf{RQ}$_2$, we assess \Model~against state-of-the-art models based on transfer learning (\texttt{BEETLE} and \texttt{tEAMS}), multi-task learning (\texttt{MORF}), and meta-learning (\texttt{MAML} and \texttt{MetaSGD}), as discussed in Section~\ref{sec:bg-transfer} and~\ref{sec:bg-meta}. Again, the same source code published by their authors is used. We also examine the approaches that take environmental features as additional inputs, i.e., \texttt{DeepPerf$+_{e}$}, \texttt{RF$+_{e}$}, and \texttt{SPLConqueror$+_{e}$} (see Section~\ref{sec:bg-env}).

      \begin{figure}[!t]
  \centering
  \begin{subfigure}[t]{0.6\columnwidth}
        \centering
\includegraphics[width=\textwidth]{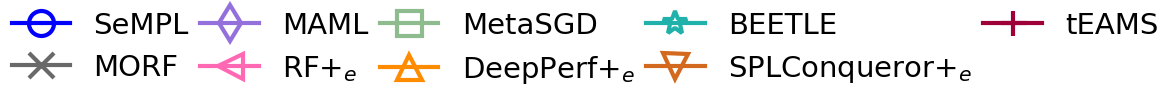}
   \end{subfigure}
%% Legend %%

   \begin{subfigure}[t]{\subfigsize\columnwidth}
        \centering
\includegraphics[width=\textwidth]{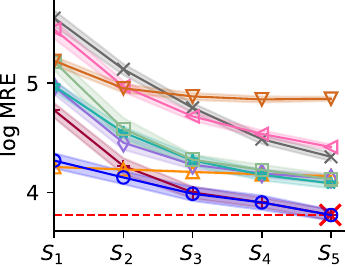}
 \subcaption*{\textsc{DeepArch}  ($sp=1.00\times$)}
   \end{subfigure}
~\hspace{-0.15cm}
\begin{subfigure}[t]{\subfigsize\columnwidth}
        \centering
\includegraphics[width=\textwidth]{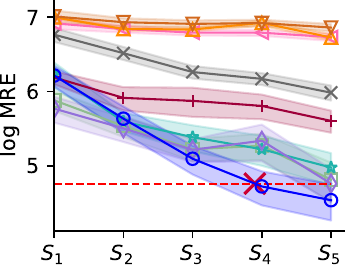}
      \subcaption*{\textsc{SaC}  ($sp=1.28\times$)}
   \end{subfigure}
~\hspace{-0.15cm}
   \begin{subfigure}[t]{\subfigsize\columnwidth}
        \centering
\includegraphics[width=\textwidth]{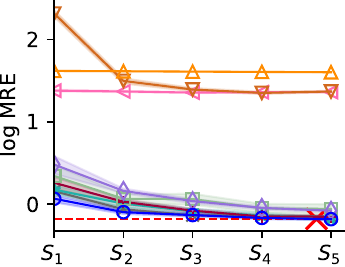}
     \subcaption*{\textsc{SQLite}  ($sp=1.04\times$)}
   \end{subfigure}
~\hspace{-0.15cm}
      \begin{subfigure}[t]{\subfigsize\columnwidth}
        \centering
\includegraphics[width=\textwidth]{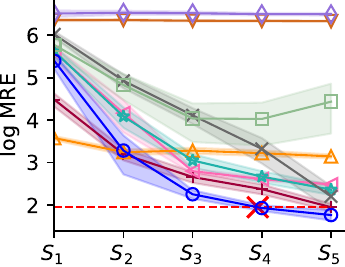}
 \subcaption*{\textsc{NGINX}  ($sp=1.27\times$)}
   \end{subfigure}
%%%%%%%%%%%%%%
   \begin{subfigure}[t]{\subfigsize\columnwidth}
        \centering
\includegraphics[width=\textwidth]{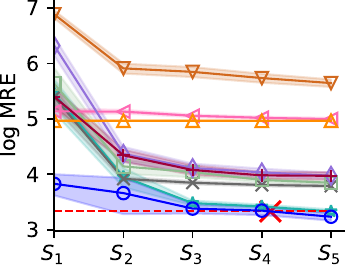}
     \subcaption*{\textsc{SPEAR}  ($sp=1.21\times$)}
   \end{subfigure}
~\hspace{-0.15cm}
    \begin{subfigure}[t]{\subfigsize\columnwidth}
        \centering
\includegraphics[width=\textwidth]{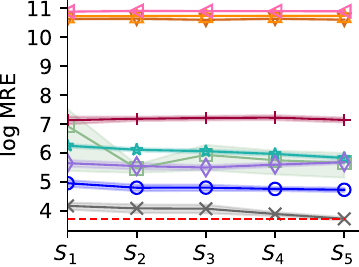}
      \subcaption*{\textsc{Storm}  ($sp=N/A$)}
   \end{subfigure}
~\hspace{-0.15cm}
      \begin{subfigure}[t]{\subfigsize\columnwidth}
        \centering
\includegraphics[width=\textwidth]{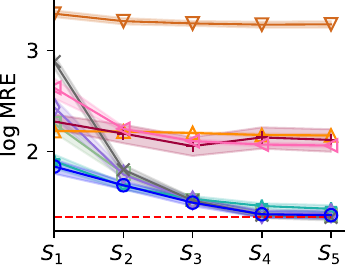}
 \subcaption*{\textsc{ImageMagick} ($sp=N/A$)}
   \end{subfigure}
~\hspace{-0.15cm}
   \begin{subfigure}[t]{\subfigsize\columnwidth}
        \centering
\includegraphics[width=\textwidth]{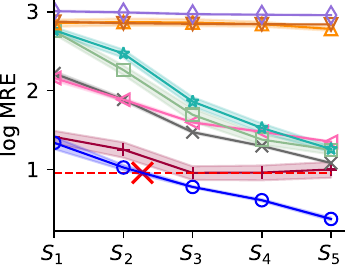}
    \subcaption*{\textsc{ExaStencils} ($sp=1.55\times$)}
   \end{subfigure}
%%%%%%%%%%%%%%

    \begin{subfigure}[t]{\subfigsize\columnwidth}
        \centering
\includegraphics[width=\textwidth]{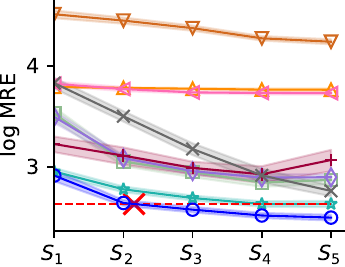}
      \subcaption*{\textsc{x264} ($sp=2.13\times$)}
   \end{subfigure}
~\hfill
 \begin{subfigure}[t]{0.67\columnwidth}
 \vspace{-2.5cm}
  %\begin{table}[!t]
%\caption{Average Scott-Knott ranks for \Model~versus multi-environment models over all cases/runs. \setlength{\fboxsep}{1.5pt}\colorbox{steel!20}{blue} is the best.}
\centering
\footnotesize
\begin{adjustbox}{width=\linewidth,center}
\begin{tabular}{l|ccccccccc}
\toprule
\textbf{Model} & \rotatebox[origin=c]{0}{\textsc{DeepArch}} & \rotatebox[origin=c]{0}{\textsc{SaC}} & \rotatebox[origin=c]{0}{\textsc{SQLite}} & \rotatebox[origin=c]{0}{\textsc{NGINX}} & \rotatebox[origin=c]{0}{\textsc{SPEAR}} & \rotatebox[origin=c]{0}{\textsc{Storm}} & \rotatebox[origin=c]{0}{\textsc{ImageMagick}} & \rotatebox[origin=c]{0}{\textsc{ExaStencils}} & \rotatebox[origin=c]{0}{\textsc{x264}}\\ 
\midrule
\Model  & \cellcolor{steel!20}\textbf{1.67}  & \cellcolor{steel!20}\textbf{1.2}  & \cellcolor{steel!20}\textbf{1.2}  & \cellcolor{steel!20}\textbf{1.55}  & \cellcolor{steel!20}\textbf{1.6}  & 2.73  & \cellcolor{steel!20}\textbf{1.9}  & \cellcolor{steel!20}\textbf{1.15}  & \cellcolor{steel!20}\textbf{1.48}  \\ 
\texttt{MAML}  & 3.53  & 2.67  & 3.95  & 7.0  & 4.6  & 3.73  & 2.8  & 7.65  & 4.16  \\ 
\texttt{MetaSGD}  & 4.2  & 1.93  & 3.4  & 4.65  & 3.87  & 3.4  & 2.5  & 4.0  & 3.9  \\ 
\texttt{BEETLE}  & 3.67  & 2.4  & 2.25  & 3.1  & 2.07  & 3.4  & 2.5  & 4.75  & 2.04  \\ 
\texttt{tEAMS}  & 1.8  & 3.47  & 2.55  & 1.85  & 4.67  & 4.87  & 4.4  & 1.85  & 4.2  \\ 
\texttt{MORF}  & 5.73  & 2.4  & 1.85  & 4.25  & 3.07  & \cellcolor{steel!20}\textbf{1.0}  & 3.25  & 2.95  & 4.78  \\ 
\texttt{RF$+_{e}$}  & 5.93  & 5.07  & 5.15  & 2.8  & 4.47  & 7.73  & 4.7  & 3.65  & 4.18  \\ 
\texttt{DeepPerf$+_{e}$}  & 2.8  & 5.8  & 6.5  & 3.0  & 4.6  & 7.07  & 3.7  & 6.3  & 4.06  \\ 
\texttt{SPLConqueror$+_{e}$}  & 6.13  & 6.4  & 6.0  & 6.0  & 7.13  & 6.4  & 6.7  & 6.4  & 7.2  \\

\bottomrule
\end{tabular}
\end{adjustbox}
%\label{tb:rank_rq3}
%\end{table}
  \vspace{-0.1cm}
\subcaption*{Mean Scott-Knott ranks over all cases/runs. \setlength{\fboxsep}{1.5pt}\colorbox{steel!20}{blue} cell denotes the best.}
 \label{fig:mappings}
    \end{subfigure}
%~\hspace{1cm}
   % \caption{The log mean MRE of the compared performance models for 5 training size and 9 subject systems in RQ$_{3}$.}
      \caption{\Model~versus multi-environment models. The format is the same as Figure~\ref{fig:rq1-mean}. Detailed MRE and ranks can be accessed at:   \texttt{\textcolor{blue}{\protect\url{https://github.com/ideas-labo/SeMPL/blob/main/Figure6_full.pdf}}}.}
      \label{fig:rq2-mean}
         \vspace{-0.4cm}
  \end{figure}

Since \texttt{MAML}, \texttt{MetaSGD}, \texttt{BEETLE}, and \texttt{tEAMS} are agnostic to the base learner, we pair them with \texttt{DeepPerf}, which is the same default for \Model. Their pre-training data is also identical to that of \Model~where applicable. All other settings are the same as \textbf{RQ$_1$}. Note that given such a setting, all the compared approaches use the same amount of data in training/fine-tuning as that of \Model, including those from the target environment.

%Though it is proved that \Model~is better than the single-task models, muli-tasks which utilize different ways of reusing knowledge from other tasks, are not yet compared. To assess \Model~against the SOTA multi-task machine learning approaches, the following models are evaluated:

%For RF$_{MTL}$, the implementation of \texttt{scikit-learn} is used in this thesis. Besides, to make a fair comparison with the transfer learning approach, this study utilizes the source selection algorithm of BEETLE and uses rDNN as the base learner. Likewise, since MAML and MetaSGD are model agnostic, rDNN is taken as the base learner for an impartial competition. The remaining of the experiment settings is the same as those used above.

\subsubsection{Results}

% \begin{table*}[]
% % \begin{sidewaystable}[]
% \footnotesize
% \renewcommand\arraystretch{1.2}
% \caption{The median and interquartile range of MRE, denoted as Med (IQR), for \Model~and the SOTA multi-task approaches for all the subject systems and training sizes over 30 runs. For each case, the one(s) with the best rank from the Scott-Knott test is highlighted in bold, and the \setlength{\fboxsep}{1.5pt}\colorbox{green!20}{\textbf{green cells}} mean \Model~is ranked the first.}
% % \begin{adjustbox}{width=\textwidth,center}
% \input{Tables/RQ3.tex}
% % \end{adjustbox}
% \label{tb:RQ3}
% \end{table*}

%The best improvement \textcolor{red}{ranges from 74\% (for \texttt{tEAMS}, \textsc{SPEAR}, $S_{1}$) to 99\% (for \texttt{MAML}, \textsc{NGINX}, $S_{4}$)

From Figure~\ref{fig:rq2-mean}, we see that in general, \Model~is mostly more accurate than, or at least similar to, the best state-of-the-art model that handles multiple environments. The best improvement ranges from 74\% (for \textsc{SPEAR}) to 99\% (for \textsc{NGINX}). In terms of data efficiency, \Model~shows considerable speedup with up to $2.13\times$. We observe similar findings on the Scott-Knott test: overall, \Model~is ranked the best for $89\%$ of the systems (8 out of 9 systems). Meanwhile, although the best rank for \textsc{Storm} is \texttt{MORF}, we can still see that \Model~ranks the second in general. 

%with only one exception on \textsc{Storm} due to the aforementioned reason.

  %\input{Tables/rank_rq3}

These results match with our theory: transfer learning models like \texttt{BEETLE} and \texttt{tEAMS} are restricted by their weak generalizability; while multi-task learning models like \texttt{MORF} overcomplicate the training, which makes the accuracy suffer; meta-learning models like \texttt{MAML} and \texttt{MetaSGD} fail to discriminate the contributions between different meta environments when pre-training the meta-model. All of the above shortcomings are what \Model~seeks to tackle. In particular, models that take the environments as additional features generally perform badly (e.g., \texttt{DeepPerf$+_{e}$}), due to the fact that this unnecessarily increases the difficulty of training. 

%and exacerbates the issue of large variation between environments.

%The results shown in Figure~\ref{fig:rq3} and Table~\ref{tb:rank_rq3} reveal that among the 9 systems, \Model~is ranked the best in 8 cases, i.e., in $88.9\%$ of the comparisons \Model~outperforms the state-of-the-arts. Moreover, for \textsc{SPEAR} and \textsc{x264}, the MRE of \Model~is better than the other performance models for all training sizes, while for the rest cases \Model~has the best ranking only except for \textsc{Storm}. To be specific, in the experiments for \textsc{SaC}, \Model~outperforms the second best artifact $37.8\%$ and $85\%$ better than the worst.

%Moreover, even if the same rDNN base learner is used for MAML, MetaSGD and BEETLE, the sequential meta model still has the lowest MRE in general. This also illustrates that training a meta model in the proper sequence can improve the accuracy. Consequently, RQ4.3 is answered as below:

%As such, we answer \textbf{RQ$_{2}$:} as:

\begin{quotebox}
   \noindent
   \textit{\textbf{RQ$_{2}$:}} Compared with the state-of-the-art models that handles multiple environments, \Model~is more effective on 89\% of the systems with the best MRE improvement from 74\% to 99\% while being data-efficient with mostly $sp \geq 1$ and the best speedup of $2.13\times$. In particular, this evidences that the sequential training in \Model, which discriminates the contributions of each meta environment, is more beneficial for configuration performance learning than the parallel training (e.g., in \texttt{MAML}) that treats all meta environments equally.
\end{quotebox}

\begin{figure}[!t]
  \centering
\begin{subfigure}[t]{\columnwidth}
        \centering
\includegraphics[width=0.85\textwidth]{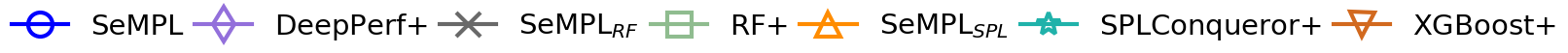}
   \end{subfigure}
%% Legend %%
  
\begin{subfigure}[t]{\subfigsize\columnwidth}
        \centering
\includegraphics[width=\textwidth]{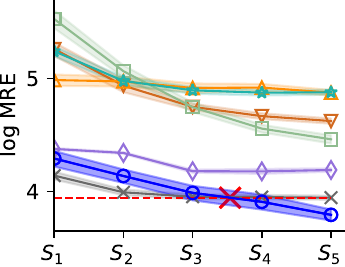}
\subcaption*{\textsc{DeepArch} ($sp=1.41\times$)}
   \end{subfigure}
~\hspace{-0.15cm}
    \begin{subfigure}[t]{\subfigsize\columnwidth}
        \centering
\includegraphics[width=\textwidth]{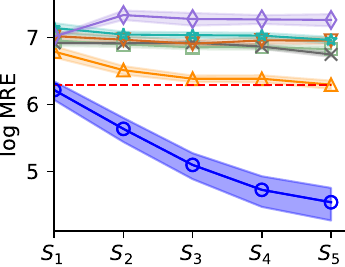}
\subcaption*{\textsc{SaC} ($sp>5.00\times$)}
   \end{subfigure}
~\hspace{-0.15cm}
   \begin{subfigure}[t]{\subfigsize\columnwidth}
        \centering
\includegraphics[width=\textwidth]{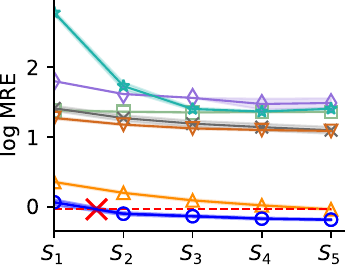}
\subcaption*{\textsc{SQLite} ($sp=3.10\times$)}
   \end{subfigure}
~\hspace{-0.15cm}
      \begin{subfigure}[t]{\subfigsize\columnwidth}
        \centering
\includegraphics[width=\textwidth]{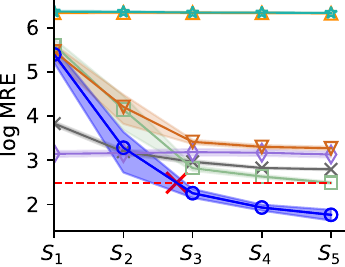}
\subcaption*{\textsc{NGINX} ($sp=1.80\times$)}
   \end{subfigure}

   \begin{subfigure}[t]{\subfigsize\columnwidth}
        \centering
\includegraphics[width=\textwidth]{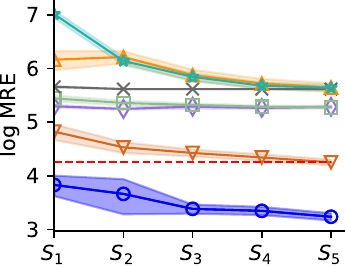}
\subcaption*{\textsc{SPEAR} ($sp>5.00\times$)}
   \end{subfigure}
~\hspace{-0.15cm}
    \begin{subfigure}[t]{\subfigsize\columnwidth}
        \centering
\includegraphics[width=\textwidth]{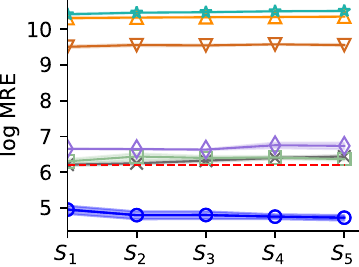}
\subcaption*{\textsc{Storm} ($sp>1.00\times$)}
   \end{subfigure}
~\hspace{-0.15cm}
      \begin{subfigure}[t]{\subfigsize\columnwidth}
        \centering
\includegraphics[width=\textwidth]{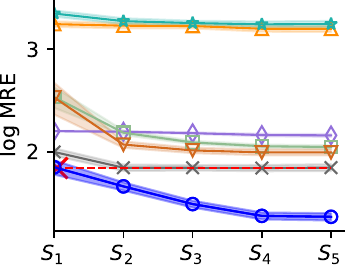}
\subcaption*{\textsc{ImageMagick} ($sp=5.72\times$)}
   \end{subfigure}
~\hspace{-0.15cm}
   \begin{subfigure}[t]{\subfigsize\columnwidth}
        \centering
\includegraphics[width=\textwidth]{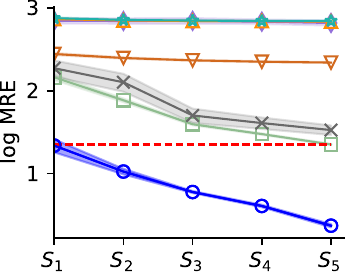}
\subcaption*{\textsc{ExaStencils} ($sp>6.56\times$)}
   \end{subfigure}

    \begin{subfigure}[t]{\subfigsize\columnwidth}
        \centering
\includegraphics[width=\textwidth]{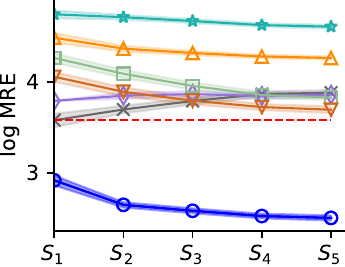}
\subcaption*{\textsc{x264} ($sp>1.00\times$)}
   \end{subfigure}
   ~\hfill
 \begin{subfigure}[t]{0.7\columnwidth}
 \vspace{-2.3cm}
  %\begin{table}[!t]
%\caption{Average Scott-Knott ranks for \Model~versus multi-environment models over all cases/runs. \setlength{\fboxsep}{1.5pt}\colorbox{steel!20}{blue} is the best.}
\centering
\footnotesize
\begin{adjustbox}{width=\linewidth,center}
\begin{tabular}{l|ccccccccc}
\toprule
\textbf{Model} & \rotatebox[origin=c]{0}{\textsc{DeepArch}} & \rotatebox[origin=c]{0}{\textsc{SaC}} & \rotatebox[origin=c]{0}{\textsc{SQLite}} & \rotatebox[origin=c]{0}{\textsc{NGINX}} & \rotatebox[origin=c]{0}{\textsc{SPEAR}} & \rotatebox[origin=c]{0}{\textsc{Storm}} & \rotatebox[origin=c]{0}{\textsc{ImageMagick}} & \rotatebox[origin=c]{0}{\textsc{ExaStencils}} & \rotatebox[origin=c]{0}{\textsc{x264}}\\ 
\midrule
% \Model  & 1.53  & \cellcolor{steel!20}\textbf{1.0}  & \cellcolor{steel!20}\textbf{1.0}  & \cellcolor{steel!20}\textbf{1.6}  & \cellcolor{steel!20}\textbf{1.73}  & \cellcolor{steel!20}\textbf{1.33}  & \cellcolor{steel!20}\textbf{1.3}  & \cellcolor{steel!20}\textbf{1.0}  & \cellcolor{steel!20}\textbf{1.18}  \\ 
% \texttt{DeepPerf$+$}  & 2.4  & 5.27  & 4.65  & 2.75  & 2.2  & 3.4  & 2.55  & 3.95  & 2.76  \\ 
% \texttt{\Model$_{RF}$}  & \cellcolor{steel!20}\textbf{1.47}  & 3.53  & 3.85  & 2.4  & 2.87  & 2.2  & 2.7  & 2.45  & 2.62  \\ 
% \texttt{RF$+$}  & 4.4  & 3.07  & 3.45  & 2.65  & 2.47  & 2.67  & 3.2  & 2.15  & 3.6  \\ 
% \texttt{\Model$_{SPL}$}  & 4.33  & 2.2  & 1.95  & 4.9  & 4.33  & 5.07  & 5.05  & 4.15  & 4.62  \\ 
% \texttt{SPLConqueror$+$}  & 4.8  & 4.73  & 4.8  & 5.4  & 4.73  & 5.4  & 5.25  & 4.35  & 5.32  \\ 

\Model  & 1.53  & \cellcolor{steel!20}\textbf{1.0}  & \cellcolor{steel!20}\textbf{1.0}  & \cellcolor{steel!20}\textbf{1.6}  & \cellcolor{steel!20}\textbf{1.73}  & \cellcolor{steel!20}\textbf{1.33}  & \cellcolor{steel!20}\textbf{1.3}  & \cellcolor{steel!20}\textbf{1.0}  & \cellcolor{steel!20}\textbf{1.18}  \\ 
\texttt{\texttt{DeepPerf}$+$}  & 2.47  & 5.8  & 5.05  & 2.85  & 2.8  & 3.4  & 3.0  & 4.85  & 2.92  \\ 
\texttt{\Model$_{RF}$}  & \cellcolor{steel!20}\textbf{1.47}  & 3.53  & 4.1  & 2.4  & 3.47  & 2.2  & 2.9  & 2.45  & 2.82  \\ 
\texttt{RF$+$}  & 5.0  & 3.07  & 3.7  & 2.65  & 3.07  & 2.67  & 3.45  & 2.15  & 3.82  \\ 
\texttt{\Model$_{SPL}$}  & 4.93  & 2.27  & 2.0  & 5.4  & 5.0  & 5.6  & 5.75  & 5.05  & 5.26  \\ 
\texttt{SPLConqueror$+$}  & 5.4  & 5.27  & 5.25  & 5.9  & 5.47  & 6.27  & 5.95  & 5.25  & 6.04  \\ 
\texttt{XGBoost$+$}  & 4.4  & 4.6  & 4.8  & 3.95  & 2.67  & 5.07  & 3.7  & 3.5  & 4.02  \\

\bottomrule
\end{tabular}
\end{adjustbox}
%\label{tb:rank_rq3}
%\end{table}
   % \vspace{0.2cm}
\subcaption*{Mean Scott-Knott ranks over all cases/runs. \setlength{\fboxsep}{1.5pt}\colorbox{steel!20}{blue} cell denotes the best.}
 \label{fig:mappings}
    \end{subfigure}
%~\hspace{1cm}
     
    %\caption{The log mean MRE of the compared performance models for 5 training size and 9 subject systems in RQ$_{2}$.}
      \caption{Optimal sequence versus random order in \Model. The format is the same as Figure~\ref{fig:rq1-mean}. Detailed MRE and ranks can be accessed at:  \texttt{\textcolor{blue}{\protect\url{https://github.com/ideas-labo/SeMPL/blob/main/Figure7_full.pdf}}}.}
      \label{fig:rq3-mean}
      \vspace{-0.5cm}
  \end{figure}

\subsection{Effectiveness of Sequence Selection}

\subsubsection{Method}

%Since in \textbf{RQ}$_1$, the pre-training in \Model~gains additional benefits from the data samples for the meta-tasks, while the others are directly trained on the target-task without pre-training. 

%To ensure that the superiority of \Model~observed for \textbf{RQ$_{1}$} and \textbf{RQ$_{2}$} is not simply due to the fact that the meta-tasks are considered sequentially, 

To confirm the necessity of sequence selection in \Model, for \textbf{RQ}$_3$, we equip the other models from \textbf{RQ}$_1$ with the sequential meta-learning process using all data of the meta environments in random orders, denoted as \texttt{DeepPerf}$+$, \texttt{RF}$+$, \texttt{SPLConqueror}$+$, and \texttt{XGBoost}$+$. Thus, they will have access to exactly the same amount of data as \Model, including those from the target enforcement for fine-tuning. Note that we cannot consider \texttt{DECART} because its mechanism requires initializing the model from scratch, which is incompatible with the concept of fine-tuning.

%for each group of training data without the ability of pre-training.

To eliminate the noise caused by the default base learner, we also additionally pair \Model~with two alternative base learners, i.e., \texttt{RF} and \texttt{SPLConqueror}, denoted as \Model$_{RF}$ and \Model$_{SPL}$, respectively. All other settings are the same as the previous RQs.

% In the previous RQ, it has been proved that \Model~is better, with the sequential pre-training process using the meta tasks. Yet, anther question to ask is, what if the single-task models are also pre-trained with the meta data? In other words, this RQ seek to know that: does the sequential training contributes to the improvements in the prediction accuracy?

% To that end, the single-task models are pre-trained using two different methods in the experiments:

% \begin{itemize}
%     \item \textbf{All-together pre-training.} First, the data of all the meta tasks are fed to the single-task models for pre-training, and then the pre-trained model is fine-tuned for the new-coming task. The final model is denoted as \texttt{model$_{PLUS}$}.
%     \item \textbf{Sequential pre-training.} Since the meta learning framework for \Model is model agnostic, the same sequential pre-training method is applied to the single-task models, and the resulting model is denoted as \texttt{model$_{SE}$}. Note that because \Model~uses DeepPerf as the base learner, DeepPerf$_{SE}$ is actually the same as \Model. Moreover, since DECART and SQLConqueror are implemented to be not pre-trainable, DECART$_{SE}$ and DECART$_{PLUS}$ are not compared in this study, similarly, SQLConqueror$_{SE}$ and SQLConqueror$_{PLUS}$ are replaced by LR$_{SE}$ and LR$_{PLUS}$ respectively.
% \end{itemize}

% In total, there are 160 groups of data in this experiment, with the 32 tasks and 5 pre-trained model to be compared. The rest of the experiment settings is the same as described in the prior subsection.

\subsubsection{Result}

% \begin{table*}[]
% % \begin{sidewaystable}[]
% \footnotesize
% \renewcommand\arraystretch{1.2}
% \caption{The median and interquartile range of MRE, denoted as Med (IQR), for \Model~and the pre-trained single-task approaches for all the subject systems and training sizes over 30 runs. For each case, the one(s) with the best rank from the Scott-Knott test is highlighted in bold, and the \setlength{\fboxsep}{1.5pt}\colorbox{green!20}{\textbf{green cells}} mean \Model~is ranked the first.}
% \begin{adjustbox}{width=\textwidth,center}
% \input{Tables/RQ2.tex}
% \end{adjustbox}
% \label{tb:RQ2}
% \end{table*}

From Figure~\ref{fig:rq3-mean}, we see that \Model, \Model$_{RF}$ and \Model$_{SPL}$ generally perform better than their counterparts on 9 (100\%), 5 (56\%), and 9 (100\%) out of 9 systems, respectively. From this, we confirm the importance of obtaining the optimal sequence for the sequential learning in \Model, as a random order would likely amplify the side-effect of some badly performing meta environments. Clearly, using \texttt{DeepPerf} as the base learner in \Model~dramatically boosts the accuracy across nearly all training sizes on 8 out of 9 systems with the best speedup of more than $6.56\times$. Note that while for \textsc{DeepArch}, \Model~has the second-best overall rank, it is very close to the best counterpart. 
% Some results (especially negative ones) are not discussed in the paper

%\input{Tables/rank_rq2}

%we observe that \Model~may worsen its accuracy with more training data when paired with \texttt{RF}, e.g., \Model$_{RF}$ for \textsc{x264}.

Interestingly, \Model$_{RF}$ might be exacerbated with more training data, e.g., for \textsc{x264}. This is because the base learner, \texttt{RF}, still suffers from the issue of overfitting under the sample sizes considered, which, when updating the meta-models over multiple meta environments, can be harmful due to the cumulatively overfitted model parameters. It is also the reason that \Model$_{RF}$ does not perform as well as \texttt{RF$+$} on some systems.

%In contrast, using \texttt{DeepPerf} as the base learner can better relieve this issue through regularization.

%

% the following key observations can be obtained: 

% \begin{itemize}
%     \item  Even pre-trained with meta data, the single-task models are still outperformed by \Model~in 8 out of 9 cases ($88.9\%$). In the best case, the rank of \Model~is $58.3\%$ lower than the best pre-trained model and $81.2\%$ better than the worst one (for the system of \textsc{SaC}).
%     \item Among the 9 systems, 7 of which \Model~has the best MRE for all the training sizes, while for 1 of the remaining systems (\textsc{NGINX}), it is the best in 4 of the sizes. In particular, we observe a degraded behaviour of pre-trained RF and \texttt{SPLConqueror} for \textsc{Storm}, possibly for the reason of the incompatibility with their optimizers.
%     \item Even for the only system that \Model~does not rank the first (\textsc{DeepArch}), it outperforms the best pre-trained model in the last 2 sample sizes and its ranking is only $4.1\%$ worse.
%     \item Worth noting, there exists an abnormal trending for some pre-trained models (\Model$_{SPL}$ for \textsc{SPEAR} and \Model$_{RF}$ for \textsc{x264}). The key reason could be the optimization algorithm of the abnormal model does not find the right direction, and therefore more samples result in worse MRE. Yet, according to our experiments, our regularized DNN can handle this issue well.
% \end{itemize}

%Overall, the answer for \textbf{RQ$_3$} is:

\begin{quotebox}
   \noindent
   \textit{\textbf{RQ$_{3}$:}} The sequence selection helps to improve the results on between 56\% and 100\% of the systems based on the base learner. Notably, pairing \Model~with \texttt{DeepPerf} achieves the best results overall and leads to considerable speedup.
\end{quotebox}

\subsection{Sensitivity to the Pre-Training Size}

\subsubsection{Method}

Instead of full pre-training sizes, in \textbf{RQ}$_4$, we examine the MRE at $0\%, 25\%,...,100\%$ of the random samples from the full datasets of meta environments, where $0\%$ means no meta-learning.

%examine the sensitivity of the accuracy to the sample size in the pre-training.

\subsubsection{Results}
%a similar patterns has been registered in the others.

%Due to space limit, we report on three randomly selected systems and cases. 

\begin{figure}[!t]
\centering
\begin{adjustbox}{width=\textwidth,center}
    \begin{subfigure}[t]{0.118\columnwidth}
    \includegraphics[width=\columnwidth]{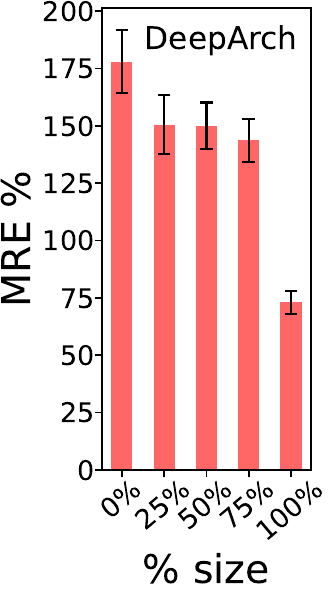}
    %\vspace{-0.2cm}
    %\subcaption{\textsc{SQlite}}
    \end{subfigure}
 ~\hspace{-0.3cm}
    \begin{subfigure}[t]{0.103\columnwidth}
    \includegraphics[width=\columnwidth]{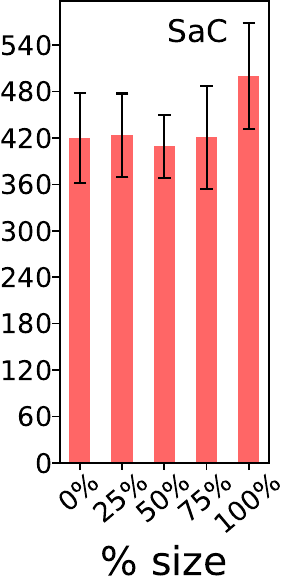}
    %\vspace{-0.2cm}
    %\subcaption{\textsc{SQlite}}
    \end{subfigure}
 ~\hspace{-0.3cm}
    \begin{subfigure}[t]{0.106\columnwidth}
    \includegraphics[width=\columnwidth]{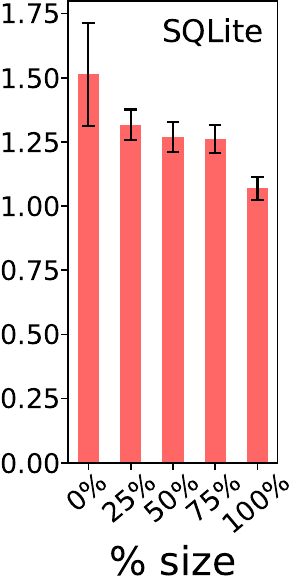}
    %\vspace{-0.2cm}
    %\subcaption{\textsc{SQlite}}
    \end{subfigure}
 ~\hspace{-0.3cm}
     \begin{subfigure}[t]{0.103\columnwidth}
    \includegraphics[width=\columnwidth]{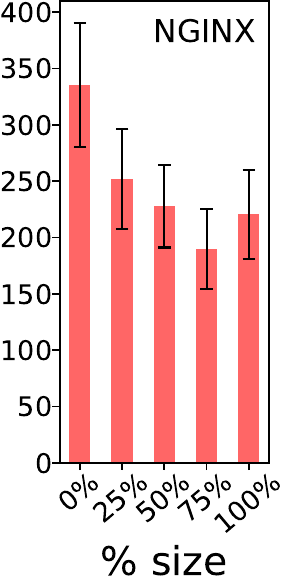}
    %\vspace{-0.2cm}
    %\subcaption{\textsc{SQlite}}
    \end{subfigure}
 ~\hspace{-0.3cm}
     \begin{subfigure}[t]{0.103\columnwidth}
    \includegraphics[width=\columnwidth]{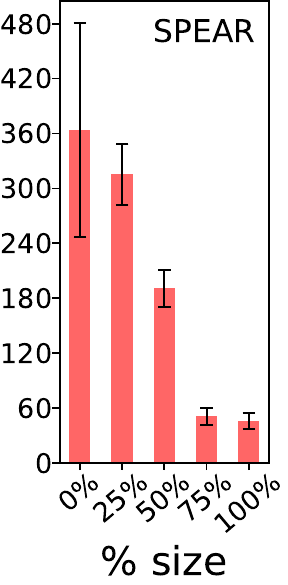}
    %\vspace{-0.2cm}
    %\subcaption{\textsc{SQlite}}
    \end{subfigure}
 ~\hspace{-0.3cm}
     \begin{subfigure}[t]{0.1027\columnwidth}
    \includegraphics[width=\columnwidth]{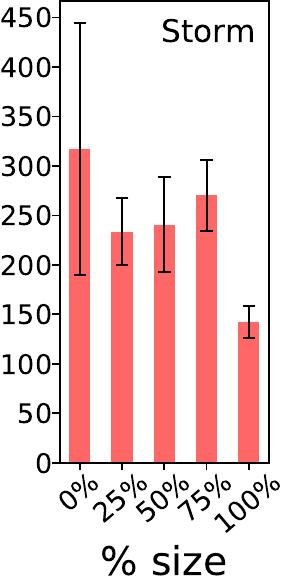}
    %\vspace{-0.2cm}
    %\subcaption{\textsc{SQlite}}
    \end{subfigure}
 ~\hspace{-0.3cm}
     \begin{subfigure}[t]{0.097\columnwidth}
    \includegraphics[width=\columnwidth]{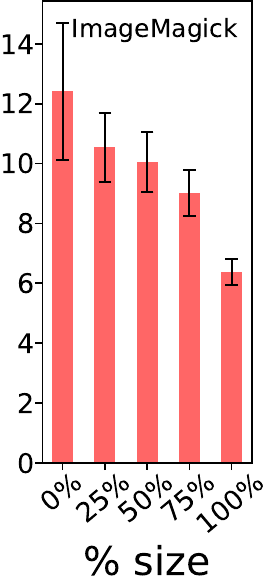}
    %\vspace{-0.2cm}
    %\subcaption{\textsc{SQlite}}
    \end{subfigure}
 ~\hspace{-0.3cm}
     \begin{subfigure}[t]{0.097\columnwidth}
    \includegraphics[width=\columnwidth]{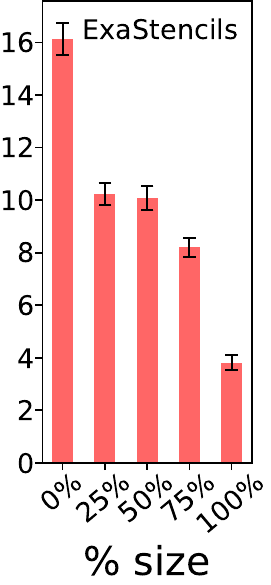}
    %\vspace{-0.2cm}
    %\subcaption{\textsc{SQlite}}
    \end{subfigure}
 ~\hspace{-0.3cm}
     \begin{subfigure}[t]{0.097\columnwidth}
    \includegraphics[width=\columnwidth]{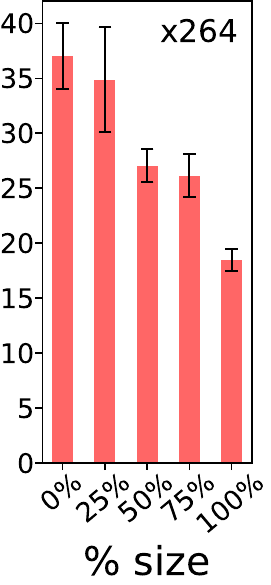}
    %\vspace{-0.2cm}
    %\subcaption{\textsc{SQlite}}
    \end{subfigure}
    
\end{adjustbox}
%\vspace{-0.2cm}
\caption{Mean MRE and standard error under different percentages of pre-training sizes over 30 runs (the sample size at the fine-tuning stage is $S_1$, which is the smallest since we seek to stress test the effect).}
\vspace{-0.3cm}
\label{fig:meta sizes}
\end{figure}

In Figure~\ref{fig:meta sizes}, we note that there is generally a monotonic correlation between the sample size of meta environments and the accuracy: the MRE decreases with more data samples used in the pre-training of \Model. In particular, the case of $0\%$ leads to the worst results on 8 out of 9 systems, suggesting the importance of meta-learning. The only exception is \textsc{SaC} which makes \Model~suffer the ``curse of dimensionality'' more since it has $58$ configuration options, whereas the others range from $5$ to $24$. Yet, the biggest improvement is case-dependent since the relative difference in contributions of the meta environments to the learning can vary across the systems.

%Although the improvement continues over $50\%$ size, the slope tends to be flattened. This means that the ability to learn from meta-tasks has already been achieved at a good level under the small size of data samples in pre-training, thanks to the sequence selection in \Model. 

%Therefore, we say:

\begin{quotebox}
   \noindent
    \textit{\textbf{RQ$_4$:}} While more data samples in the pre-training can be generally beneficial to \Model~but the biggest improvement is case-dependent. 
   %\textit{\textbf{RQ$_4$:}} More samples in pre-training can indeed improve the accuracy of \Model~on the target-task, but the convergence rate is acceptable and the biggest improvement occurs as soon as the meta learning is used (from $0\%$). 
\end{quotebox}

% \begin{figure}[!t]
% \centering
%     \begin{subfigure}[t]{0.26\columnwidth}
%     \includegraphics[width=\columnwidth]{Figures/RQ4/case1.pdf}
%     %\vspace{-0.2cm}
%     %\subcaption{\textsc{SQlite}}
%     \end{subfigure}
%  ~\hspace{0.5cm}
%     \begin{subfigure}[t]{0.25\columnwidth}
%     \includegraphics[width=\columnwidth]{Figures/RQ4/case2.pdf}
%     %\vspace{-0.2cm}
%     %\subcaption{\textsc{IM}}
%     \end{subfigure}
%      ~\hspace{0.5cm}
%     \begin{subfigure}[t]{0.25\columnwidth}
%     \includegraphics[width=\columnwidth]{Figures/RQ4/case3.pdf}
%     %\vspace{-0.2cm}
%     %\subcaption{\textsc{IM}}
%     \end{subfigure}
% %\vspace{-0.2cm}
% \caption{Mean MRE and standard error under different percentages of pre-training sample sizes over 30 runs.}
% \vspace{-0.4cm}
% \label{fig:meta sizes}
% \end{figure}

% \begin{figure}[!t]
% \centering
%     \begin{subfigure}[t]{0.51\columnwidth}
%     \includegraphics[width=\columnwidth]{Figures/RQ4/sqlite-overwritebatch-4tasks_meta_sizes_mean.png}
%     %\vspace{-0.2cm}
%     \subcaption{\textsc{SQlite}}
%     \label{subfig:deeparch_meta_sizes}
%     \end{subfigure}
% % ~\hspace{-0.4cm}
%     \begin{subfigure}[t]{0.47\columnwidth}
%     \includegraphics[width=\columnwidth]{Figures/imagemagick-4tasks_meta_sizes_mean.png}
%     %\vspace{-0.2cm}
%     \subcaption{\textsc{IM}}
%     \label{subfig:imagemagick_meta_sizes}
%     \end{subfigure}
% \vspace{-0.2cm}
% \caption{Mean MRE of different pre-training sizes over 30 runs.}
% \vspace{-0.4cm}
% \label{fig:meta sizes}
% \end{figure}

% \input{discussion}
% \input{related}
\section{Discussion}
\label{sec:discussion}

%\subsection{Can \Model~Eliminate the Effect of Data From Misleading Environments?}

\subsection{What Make \Model~Special for Configuration Performance Learning?}

The proposed sequential pre-training is a specialization of meta-learning for the unique characteristics of configuration data between different environments (as discussed in Section~\ref{sec:premises}). As such, \Model~is specifically designed based on our understanding of the software engineering task at hand.

Indeed, we do not claim that \Model~can completely eliminate the potentially misleading meta environments, nor it is feasible to do so since this may incur significant overhead. However, with the sequential pre-training, \Model~is able to at least mitigate the effect of those misleading meta environments by placing them earlier in the sequence. This, as we have shown, can obtain significant improvement in accuracy over classic frameworks like \texttt{MAML}, which treats all the meta environments equally. As we discussed in Section~\ref{sec:premises}, even though certain meta environments may be misleading overall, some of their parameter distribution could still fit well with those needed for the target environment. Hence, mitigating their effect could be more suitable than their complete omission.

\subsection{Practical Application and Overhead of \Model}

\Model~was designed under the assumption that there are available data measured under different environments of a running configurable software system, which is not uncommon~\cite{DBLP:conf/sigsoft/JamshidiVKS18, DBLP:journals/corr/abs-1911-01817, DBLP:conf/icse/webertwins, LESOIL2023111671}. Therefore, the applications of \Model~in real-world cases are straightforward. Thanks to the sequential pre-training, the contributions of meta environments can be discriminated with no manual selection required, hence the more useful ones will contribute more to the learning. Of course, like any learning models, software engineers would need to decide on the training sample size to be measured for the target environment, together with whether the default base learner in \Model, i.e., \texttt{DeepPerf} that favors accuracy, is suitable for the specific scenario as the training of the base learner would constitute the majority of the \Model's overhead during pre-training.  

%The overhead of sequence selection in \Model~is also acceptable, thanks to the linear regression. 

On a machine with CPU 2GHz and 16GB RAM, \Model~only needs 2 and 60 seconds to find the optimal sequence of learning meta environments on a system with 3 and 10 meta environments, respectively, thanks to the linear regression. Of course, the overhead can increase with more meta environments, but the magnitude is still acceptable since the pre-training is an offline process.

%TODO compare with the overhead of parallel training?

\subsection{Threats to Validity}
\label{sec:threats}

{\textbf{Internal threats}} is concerned with the parameter settings, e.g., the training size. We mitigate this by following the same settings used in the state-of-the-art studies~\cite{DBLP:journals/sqj/SiegmundRKKAS12, DBLP:conf/kbse/GuoCASW13, DBLP:conf/icse/HaZ19, DBLP:conf/esem/ShuS0X20}. We also examined the sensitivity of \Model~to the pre-training sample size in RQ$_4$. \Model~is tested with \texttt{DeepPerf} as the base learner, which can be flexibly replaced. Of course, as in RQ$_3$, replacing \texttt{DeepPerf} could lead to different results but this is up to the software engineers to decide if other factors, such as training efficiency, are more important. The use of linear regression as a surrogate of the base learner might also raise threats to internal validity since it might not be perfectly accurate. However, our results show this still leads to significant improvement compared with the state-of-the-art.

%{\textbf{Internal Threats.}} Internal validity is concerned with the model parameters, e.g., the training size. We mitigate the internal threats by conducting a large amount of empirical studies on choosing the parameters, disclosing three key properties as explained in Section~\ref{sec:premises}. We also examined specifically the sensitivity of \Model~to the meta training sample size in RQ4. Moreover, the same experimental settings and evaluation methods in state-of-the-art studies~\cite{DBLP:journals/sqj/SiegmundRKKAS12,DBLP:conf/kbse/GuoCASW13,DBLP:conf/icse/HaZ19,DBLP:conf/seem/ShuS0X20} are used to strengthen the internal validity.

{\textbf{Construct threats}} are mainly related to the evaluation metric. In this work, we use MRE as the main accuracy metric as it is less sensitive to the scales in different environments and systems while being the most common metric from the literature of configuration performance learning~\cite{DBLP:conf/icse/HaZ19, DBLP:conf/esem/ShuS0X20, DBLP:journals/ese/GuoYSASVCWY18}. To measure the speedup, we follow the same protocol as used in prior work~\cite{DBLP:conf/icse/0003XC021,DBLP:conf/sigsoft/0001L21} from the field.

{\textbf{External threats}} could lie in the systems and environments studied. To mitigate such, we exploit the most commonly used datasets for diverse systems~\cite{DBLP:journals/corr/abs-1911-01817, DBLP:conf/sigsoft/JamshidiVKS18, LESOIL2023111671}. For each of these, we select all the eligible environments for evaluation. The experiments are also repeated 30 times, which tends to be sufficient to reduce randomness. The Scott-Knott test is adopted to ensure the statistical significance of the results. However, we acknowledge that using more systems and environments might be beneficial.

\section{Conclusion}
\label{sec:conclusion}

To deal with multiple environments when predicting performance for configurable software systems, this paper proposes a new category of framework that leverages meta-learning, dubbed \Model. What makes \Model~unique is that it learns the meta environments, one at a time, in a specific order according to the likely usefulness of the meta environment to the unforeseen target environment. Such sequential learning, unlike the existing parallel learning, naturally allows the pre-training to discriminate the contributions between meta environments, thereby handling the largely deviated samples from different environments---a key characteristic of the configuration data. Extensive experiments on nine widely-used systems in current studies demonstrate that \Model~significantly outperforms state-of-the-art single/multi-environment models with considerable speedup. As such, \Model~is more useful for predicting configuration performance in new environments.

%Compared with 15 state-of-the-art models on 9 systems, 3-10 meta-tasks, and 5 training sizes, we 

% \begin{itemize}
%     \item significantly outperforms the single/multi-environment models with considerable speedup;
%     \item can be greatly benefited from the sequence selection;
%     \item can be improved quickly as long as some data is available for the pre-training. 
% \end{itemize}

In the future, we hope that the concept behind \Model~will inspire a vast direction of research for meta-performance learning, including heterogeneous meta-learning and more precise management of the sample variations between environments in configuration data.

\section*{Data Availability}

All source code and data can be found at our repository: \url{https://github.com/ideas-labo/SeMPL}.

\begin{acks}
This work was supported by a UKRI Grant (10054084) and a NSFC Grant (62372084).
\end{acks}

\balance
\bibliographystyle{ACM-Reference-Format}
\bibliography{references}

\end{document}